\documentclass[11pt, draftcls, onecolumn]{IEEEtran}

\usepackage{color,epsf,psfrag,amssymb,amsfonts,latexsym,cite,verbatim}
\usepackage{enumerate,times,array, float}
\usepackage{amsmath,cases,graphicx}
\usepackage[mathscr]{eucal}

\def\psfancypar#1#2{\begingroup\def\par{\endgraf\endgroup\lineskiplimit=0pt}
               \setbox2=\hbox{\large\sc #2}
               \newdimen\tmpht \tmpht \ht2 \advance\tmpht by \baselineskip
               \font\hhuge=Times-Bold at \tmpht
               \setbox1=\hbox{{\hhuge #1}}
               \count7=\tmpht \count8=\ht1
               \divide\count8 by 1000 \divide\count7 by \count8 
               \tmpht=.001\tmpht\multiply\tmpht by \count7 
               \font\hhuge=Times-Bold at \tmpht
               \setbox1=\hbox{{\hhuge #1}}
               \noindent
                \hangindent1.05\wd1
               \hangafter=-2 {\hskip-\hangindent
               \lower1\ht1\hbox{\raise1.0\ht2\copy1}%
                \kern-0\wd1}\copy2\lineskiplimit=-1000pt}

\def\Upsilonbf{\hbox{\boldmath$\Upsilon$\unboldmath}}
\newcommand{\Phibf}{\mbox{${\bf \Phi}$}}
\newcommand{\Gammabf}{\mbox{${\bf \Gamma}$}}
\newcommand{\Psibf}{\mbox{${\bf \Psi}$}}

\newcommand{\etabf}{\mbox{${\bf\eta}$}}

\newcommand{\E}{\mbox{{\rm E}}}

\newcommand{\abf}{\mbox{${\bf a}$}}

\def\boxit#1{\vbox{\hrule\hbox{\vrule\kern3pt
        \vbox{\kern3pt#1\kern3pt}\kern3pt\vrule}\hrule}}

\def\reals{ { {\rm  I \kern-0.15em R }  } }
\def\complex{ {\,{{\rm C} \kern-0.50em \raise0.20ex {  |}}\, }}

\def\etabf{\hbox{\boldmath$\eta$\unboldmath}}

\def\mubf{\hbox{\boldmath$\mu$\unboldmath}}

\def\Sigmabf{\hbox{$\bf \Sigma$}}
\def\Upsilonbf{\hbox{$\bf \Upsilon$}}

\def\Deltabf{\hbox{$\bf \Delta$}}
\def\Gammabf{\hbox{$\bf \Gamma$}}

\def\Lambdabf{\mbox{$ \bf \Lambda $}}

\def\Pibf{{\bf \Pi}}
\def\abf{{\bf a}}

\def\dbf{{\bf d}}

\def\gbf{{\bf g}}

\def\nbf{{\bf n}}

\def\rbf{{\bf r}}
\def\sbf{{\bf s}}

\def\ubf{{\bf u}}
\def\vbf{{\bf v}}
\def\wbf{{\bf w}}
\def\xbf{{\bf x}}
\def\ybf{{\bf y}}

\def\rbf{{\bf r}}
\def\xbf{{\bf x}}
\def\ybf{{\bf y}}
\def\Abf{{\bf A}}
\def\Bbf{{\bf B}}
\def\Cbf{{\bf C}}
\def\Dbf{{\bf D}}

\def\Hbf{{\bf H}}
\def\Ibf{{\bf I}}

\def\Mbf{{\bf M}}
\def\Nbf{{\bf N}}

\def\Pbf{{\bf P}}
\def\Qbf{{\bf Q}}
\def\Rbf{{\bf R}}

\def\Ubf{{\bf U}}
\def\Vbf{{\bf V}}

\def\Xbf{{\bf X}}

\def\Cc{{\cal C}}

\def\Hc{{\cal H}}

\def\Lc{{\cal L}}
\def\Mc{{\cal M}}

\def\Qc{{\cal Q}}
\def\Rc{{\cal R}}
\def\Sc{{\cal S}}

\def\be{\vskip .3cm \begin{equation}}
\def\ee{\end{equation} \vskip .4cm \noindent}
\newcommand{\R}{\mbox{$\hat {\bf R}_{N}$}}

\def\Rxx{\Rbf_{\ssstyle X\kern-.1em X}}

\let\ssstyle=\scriptscriptstyle

\def\Kout{\setbox1=\hbox{\Huge\bf K}\hbox to
1.05\wd1{\hspace{.05\wd1}
}}
\def\Sout{\setbox1=\hbox{\Huge\bf S}\hbox to 1.05\wd1{\hspace{.05\wd1}
}}

  \ifx\LabelFigloaded\MYundefined\relax
  \else
   \fi

  \def\LabelFigloaded{\relax}

  \chardef\LabelFigCatAt\the\catcode`\@
  \catcode`\@=11

 \let\LabelFigwlog@ld\wlog
 \def\wlog#1{\relax}

 \ifx\\\MYundefined@
    \let\\\relax
 \fi

  \def\ms@g{\immediate\write16}

 \def\N@wif{\csname newif\endcsname }
 \def\Temp@ {\N@wif\ifIN@}
 \ifx\INN@\MYundefined@
    \else \let\Temp@\relax
 \fi
 \Temp@

  \def\IN@{\expandafter\INN@\expandafter}
  \long\def\INN@0#1@#2@{\long\def\NI@##1#1##2##3\ENDNI@
    {\ifx\m@rker##2\IN@false\else\IN@true\fi}%
     \expandafter\NI@#2@@#1\m@rker\ENDNI@}
  \def\m@rker{\m@@rker}
 
  \newtoks\Initialtoks@  \newtoks\Terminaltoks@
  \def\SPLIT@{\expandafter\SPLITT@\expandafter}
  \def\SPLITT@0#1@#2@{\def\TTILPS@##1#1##2@{%
     \Initialtoks@{##1}\Terminaltoks@{##2}}\expandafter\TTILPS@#2@}

 \def\Shifted@@#1#2#3{\setbox0=\hbox{#3}%
   \raise -\dp0\vbox {\kern-#2%
       \hbox {\kern#1\unhbox0\kern-#1}%
           \kern#2}}

 \newcount\gridcount
 \newbox\auxGridbox@ \newbox\hGridbox@ \newbox\vGridbox@
 \newbox\Labelbox@ \newbox\auxLabelbox@
 \newbox\Coordinatebox@
 \newtoks\Labeltoks@
 \newdimen\Wdd@ \newdimen\Htt@
 \newdimen\Wddd@ \newdimen\Httt@
 
 \def\Wr@{\immediate\write16}

 \newdimen\GL@wd
 \def\GridLineWidth#1{\GL@wd=#1}

 \def\gobble#1{}
 \def\EdgeErr@{\Wr@{}%
      \Wr@{\string\Edges\space argument
      1, 10, 100 or 1000 please\string!}%
      }

 \newcount\Edgect@

 \def\Sweepup#1\endSweepup{}

 \def\SetEdges@{%
    \edef\Zr@@s{\expandafter\gobble\number\Edgect@\empty}%
        \count255=0\Zr@@s\relax
        \ifnum\count255=\z@\else\EdgeErr@\show\tailtest\fi
        \count255=1\Zr@@s\relax
        \ifnum\count255=\Edgect@\relax\else\EdgeErr@\show\leadtest\fi
    \EdgGl@b\edef\Zr@s{\expandafter\gobble\Zr@@s\empty}
    \ifnum\Edgect@>\@ne\relax\EdgGl@b\let\L@Dc\empty
        \else\EdgGl@b\edef\L@Dc{\string.}\fi
    \ifnum\Edgect@>\@ne\relax
        \EdgGl@b\edef\Edgescale@##1{\divide##1 by \Edgect@}%
        \else\EdgGl@b\edef\Edgescale@##1{}\fi
    }

 \def\Edges#1{\Edgect@=#1\relax
     \let\EdgGl@b\global \SetEdges@}

 \Edges{1}

 \def\hhrule{\hrule height \GL@wd\vskip-.\GL@wd}

 \def\hRule@{%
   \advance\gridcount -2%
   \vfil\hhrule\vfil
   \llap{\smash{\raise -2.5pt
     \hbox{\L@Dc\number\gridcount\Zr@s\kern2pt}}}%
   \hhrule
   }

\def\vvrule{\vrule width \GL@wd \kern-\GL@wd}

 \def\vRule@{\advance\gridcount 2%
   \hfil\vvrule\hfil
   \setbox\auxGridbox@=\vbox to 0pt
      {\vskip \Htt@\vskip 2pt
        \hbox to 0pt{\hss\L@Dc\number\gridcount\Zr@s\hss}\vss}%
      \wd\auxGridbox@=0pt \box\auxGridbox@
   \vvrule
   }

 \def\PlaceGrid@@{\gridcount=10 
  \setbox\hGridbox@=\hbox{%
        \hbox{%
             \hskip-.4pt\vrule
             \vbox to \Htt@{%
               \offinterlineskip\parindent=\z@\relax
               \hbox to \Wdd@{\hfil}
               \hRule@\hRule@\hRule@\hRule@
               \vfil\hhrule\vfil}%
             \vrule\hskip-.4pt}
    }%
  \gridcount=0%
  \setbox\vGridbox@=\hbox{%
      \vbox{\offinterlineskip\parindent=0pt\hsize=0pt
         \vskip-.4pt\hrule%
         \hbox to \Wdd@{%
                 \vtop to \Htt@{\vfil}%
                 \vRule@\vRule@\vRule@\vRule@
                 \hfil\vvrule\hfil}%
         \hrule\vskip-.4pt}}%
  \wd\hGridbox@=0pt\ht\hGridbox@=0pt
  \wd\vGridbox@=0pt\ht\vGridbox@=0pt
  \hbox{\box\hGridbox@\box\vGridbox@}%
  }

 \def\LabelsGlobal{\def\LabGl@b{\global}}
 \def\LabelsLocal{\def\LabGl@b{}}
 \LabelsGlobal 

 \def\SetLabels#1\endSetLabels{%
   \LabGl@b\Labeltoks@={#1()\\}%
   }

 \LabGl@b\Labeltoks@={()\\}

 \def\ShowGrid{\LabGl@b\let\PlaceGrid@\PlaceGrid@@}
 \def\HideGrid{\LabGl@b\let\PlaceGrid@\relax}
 \def\Grids{\ShowGrid\LabGl@b\let\GridSwitch@\ShowGrid}
 \def\noGrids{\HideGrid\LabGl@b\let\GridSwitch@\HideGrid}

 \noGrids

 \def\bAdjust@@{%
     \setbox\auxLabelbox@=\hbox{\raise \dp\auxLabelbox@
            \box\auxLabelbox@}}
 \def\bAdjust@{\let\vAdjust@\bAdjust@@}

 \def\eAdjust@@{\dimen0=-.5\ht\auxLabelbox@
     \advance\dimen0 by .5\dp\auxLabelbox@
     \setbox\auxLabelbox@=
            \hbox{\raise\dimen0\box\auxLabelbox@}}
 \def\eAdjust@{\let\vAdjust@\eAdjust@@}

 \def\tAdjust@@{%
     \setbox\auxLabelbox@=\hbox{\raise-\ht\auxLabelbox@
            \box\auxLabelbox@}}
 \def\tAdjust@{\let\vAdjust@\tAdjust@@}

 \let\vAdjust@\relax

 \def\lAdjust@{\let\hAdjust@\rlap}
 \def\rAdjust@{\let\hAdjust@\llap}

 \let\hAdjust@\relax\let\vAdjust@\relax

 \def\FetchLabel@#1(#2)#3\\{%
     \IN@0#2@@\ifIN@
        \setbox0=\hbox{\ignorespaces#1#3\unskip}%
        \ifdim\wd0>0pt
           \ms@g{}%
           \ms@g{ !!! Bad label(s)? !!!}%
           \message{ #1(#2)#3}%
        \fi
        \def\LabelMole@##1\endFetchLabel@{%
            \IN@0()\\@##1@%
            \ifIN@\def\Temp@{\FetchLabel@##1\endFetchLabel@}%
            \else\def\Temp@{}%
            \fi
            \Temp@
           }%
     \else
       \ignorespaces#1\unskip
       \setbox\auxLabelbox@=%
         \hbox to 0pt{\hss\ignorespaces\hAdjust@
          {\ignorespaces#3\unskip}\hss}%
       \vAdjust@
       \let\hAdjust@\relax\let\vAdjust@\relax
       \AugmentLabelBox@@{#2}%
       \ht\Labelbox@=0pt\dp\Labelbox@=0pt
       \let\LabelMole@\FetchLabel@%
     \fi\LabelMole@}

 \newtoks\XYSep@ 
 \def\SetXYSeparator#1{%
     \IN@0#1@@\ifIN@\XYSep@{*}%
     \else
     \XYSep@{#1}%
     \fi
     }

 \SetXYSeparator*

 \def\AugmentLabelBox@@#1{%
     \IN@0\the\XYSep@ @#1@\ifIN@
       \SPLIT@0\the\XYSep@ @#1@%
       \setbox\Labelbox@=\hbox to 0pt{%
         \unhbox\Labelbox@
         \Shifted@@{\the\Initialtoks@\Wddd@}%
         {\the\Terminaltoks@\Httt@}%
         {\box\auxLabelbox@}}%
     \else
         \ms@g{}%
         \ms@g{ !!! Bad insertion point. !!!}%
         \message{ (#1\ this point was rejected.)}%
     \fi
    }

 \def\FetchOption@#1[#2]#3\endFetchOption@{%
    \def\temp{#1}
    \ifx\temp\empty
       \Edgect@=#2\relax
       \let\EdgGl@b\relax
       \SetEdges@
       \Cleaner@#3%
    \fi}

 \def\Cleaner@#1[@]{\Labeltoks@{#1}}
     
 \def\PlaceLabels@@{\mathsurround=0pt
     \def\Cr@{\\}%
     \let\L\lAdjust@\let\R\rAdjust@
     \let\B\bAdjust@\let\E\eAdjust@\let\T\tAdjust@
     \expandafter\FetchOption@\the\Labeltoks@[@]\endFetchOption@
     \Wddd@=\Wdd@ \Edgescale@\Wddd@ 
     \Httt@=\Htt@ \Edgescale@\Httt@
     \expandafter\FetchLabel@\the\Labeltoks@\endFetchLabel@
     \box\Labelbox@
     }%

 \let \PlaceLabels@\PlaceLabels@@

 \def\AffixLabels#1{\setbox\Coordinatebox@=\hbox{#1}%
      \Wdd@=\wd\Coordinatebox@ \Htt@=\ht\Coordinatebox@
      \advance\Htt@ \dp\Coordinatebox@
      \hbox{\copy\Coordinatebox@\kern-\Wdd@ 
           \Shifted@@{0pt}{-\dp\Coordinatebox@}%
           {\PlaceLabels@\PlaceGrid@}%
           \kern\Wdd@}%
      \GridSwitch@ 
      \LabGl@b\Labeltoks@{()\\}%
      }
 
   \let\wlog\LabelFigwlog@ld   
   \catcode`\@=\LabelFigCatAt  

                                By

              Raymond S\'eroul <A18645@FRCCSC21.BITNET>
                                and 
              Laurent Siebenmann <lcs@topo.math.u-psud.fr>
    
              VERSIONS: July 1991, Oct 1991, Jan 1992, July 1992

INTRODUCTION

      This labelling package is intended for TeX users who
rely on non-TeX sources for for their graphics inserts.  It
provides means for adding TeX labels to such inserts with a
minimum of fuss. 

       For most labels, TeX users have in the past found it
reasonably convenient to rely on non-TeX sources. Typical
occasions when an inescapable need for TeX labels seemed to
arise are

 (a) when the graphics program lacks certain exotic or complex
mathematical symbols

 (b) when the very highest typographical quality is wanted for the
labels

 (c) when labels included with the graphics fail to print, 
 and you cannot figure out why (cf. boxedeps.doc).  The labels
 provided by labelfig.tex are 100

       Since this package first appeared, many users, who in the
past scarcely dreamed of using TeX labels, have come to use
nothing but.  So it is now appropriate to add

Intoxication Warning:  TeX labels may be addictive and expensive. 

     If you have a fast preview you may disagree, and even find
that this package provides an agreeable paste-up environment; see
extra applications at end.

     Note to publishers: It is possible and convenient to ultimately
export the TeX labels produced by labelfig.tex to become an integral
part of the EPS file. This is often desired by a publisher who typically
uses an "upmarket" graphics or page layout program, with which the
staff is skilled in perfecting figures.  See Appendix I for
a recipe.

     The authors are grateful to Patrick Ion of Math Reviews for
helpful comments and encouragement.

BASIC INSTRUCTIONS

    After reading in the macro file using

preview or proof your figure with a coordinate grid printed on
top, by typing the following:

    \ShowGrid  
    \AffixLabels{<the graphics insertion>}

Here <the graphics insertion> is what you would type to insert
the graphics object alone without the grid.  This must provide
for the space around it. For example <the graphics insertion>
might well be \BoxedEPSF{MyFigure scaled 700} using the
boxedeps.tex macro package (from same source); this provides a
TeX box containing the encapsulated PostScript insert specified by
the file MyFigure. \AffixLabels{...} provides the grid (supposing
\ShowGrid is present) and later, once you have specified labels
using the grid, it will "tack on" the labels.

     The grid is a sort of (usually elongated) checkerboard of
ten rows and ten columns and its (internal) partitions are by
default numbered  .1, ... ,.9  both horizontally (X-coordinate
running left to right) and vertically (Y-coordinate running bottom
to top).  Thus the points enclosed by the grid correspond to the
points of the unit square in the cartesian "X-Y" plane, the lower
left corner corresponding to the origin (0,0).  By extrapolation,
the full page corresponds to a larger rectangle in the plane.

     These coordinates serve to position labels as follows.
Before the \AffixLabels{...} command type label specifications:

  \SetLabels
   (<X-coordinate>*<Y-coordinate>) <first label> \\
   .
   .
   .
   (<X-coordinate>*<Y-coordinate>)  <last label> \\
  \endSetLabels

Each row specifies one label and is terminated by \\.  In each
row, the position indicator comes first; it is written as a
standard cartesian point except that the X- and Y- coordinates
are separated by * rather than a comma because TeX allows a
comma as decimal point. There are no dimension units to specify
as the unit is the grid itself.

     By default, this cartesian point specifies where the middle
of the baseline of the label will be located.  However if you precede
the point by \L [or \R] the left [or right] edge of the baseline will
be located there. Similarly you may also precede the point by \T, \E,
or \B to vertically align the top equator or bottom of the label box
at the specified point.  This gives nine standard positions of
the label with respect to the insertion point --- corresponding to
the eight principle points of the compas and the center

                     \L\T     \T      \R\T

                     \L\E     \E      \R\E

                     \L\B     \B      \R\B

But this neglects the default "baseline" level of TeX,
giving potentially three more positions

                     \L    <no tag>   \R

For text, the baseline level is often the preferred. Its relation to
the others is variable. It will often coincide with the bottom level,
as happens for "X".  But it is often distinct, as for "g", in which
case you have in all 12 distinct positions rather than 9.

     It is convenient to think of this specification of label
position as attaching the label by a thumb-tack to the coordinate
grid. There are up to twelve positions of the thumb-tack on the
label, while the position of the thumb-tack on the coordinate grid is
arbitrary.  Normally, one choses the position of the thumb-tack on
the label to be the one that is the closest to the item being
labeled.  There are good reasons for this "rule of thumb":

   (a)  It facilitates correct positioning at first try.

   (b)  If the scale of the figure must be altered after labels
have been affixed, the labels have a good chance of remaining well
positioned.

   (c)  The visible grid need not extend beyond the "bounding box"
for the figure, because the best preferred position is always
(at least almost) within the bounding box .

The second reason is particularly important. Indeed it often
happens that scale has to be altered after labelling begins, in
order to either provide space for the labels, or to adjust
proportions between the labels and the figure.  (The size of labels
is unaffected by scaling.)

     Here is an artificial but self-contained test which uses
TeX rules to make a graphics object.

TEST

    Do not skip this!

 \def\FrameIt#1{\hbox{\vrule$\vcenter {\hrule\kern3pt%
             \hbox {\kern3pt #1\kern3pt}%
               \kern3pt\hrule}$\relax\vrule}}

 \def\Caption#1#2{\FrameIt{%
       \vtop {\hsize=#1\relax \parindent=0pt
         \leftskip=0pt \rightskip=0pt plus15pt
         \parfillskip=0pt
         \lineskip=1pt\baselineskip=0pt
         #2}}}

 \def\FirstQuadrant{\hbox to 100pt{\vrule\vbox to 100pt{%
        \hbox to 100pt{\hfil}\vfil\hrule}\hss}}

  \SetLabels
    \R(.5*.2) $\zeta\,\cdot$\\
    (.9*-.10) $\xi$\\
    \R(-.03*.9) $\eta$\\
    \T(.5*.9) \Caption{70pt}{%
          \it The norm of
          $g(\xi+i\eta)$ is indicated on
          contours of this invisible surface.}\\
  \endSetLabels

  \AffixLabels{\FirstQuadrant}

  \end

  Note that the coordinates to use for labels are indicated on the
edges of the grid (when visible) corresponding to the conventional
x- and y- axes of the Cartesian plane. By default the grid is
1-by-1. However, by the command \Edges{100}, you can change this
to 100-by-100 and many users find this alternative most
convenient. Place the command \Edges{...} in your style file (or
header) since its effect is is global. Other possible edge values
are 10 and 1000.

  If you use the command \Edges{...} at all, do so with care.  For
if you accidentally delete an \Edges{...} command your labels will
abruptly be badly misplaced and may logically but mysteriously
generate "dimension too big" errors under TeX and "off page" errors
under your driver.  

  You can dictate the edgescale for an individual figure by giving
the scale in brackets immediately after \SetLabels.  Thus, to
import into an article using say \Edge{100} a figure labelled using
another edgescale, say the original 1-by-1 default, you can use
\SetLabels[1]...\endSetLabels.

GETTING IT DOWN PAT

     Complicated labeling deserves the same respect as
complicated mathematics.  Do not expect it to come out perfect the
first time!  What is needed in either case is a mechanism to
repeatedly typeset troublesome pieces.

     One mechanism is always available.  One does complicated
labelling in a separate "test" file involving just the figure being
labelled;  a texpert will know how to \dump TeX's current state as
a temporary format that restarts rapidly at each retry.  Usually,
one then pastes the completed labelled figure back into the main
TeX file, but, of course, one can also \input it as an auxiliary
file.

     If you do not have a TeXpert at handy, here is a first
approximation to an efficient setup. By deletions reduce a copy
of your article to just a few lines before and after the figure.
Now label the figure, and finally, copy and paste the labelled
figure to the original article. Then copy the next figure to label
into this testbed and repeat. The TeXpert can improve the  speed
at which TeX starts up, by compiling a format specifically for
your article; just one caution: best NOT include in the format
ephemeral details of setup like \Set<mydriver>ArtSpecials (from
boxedeps.tex because this reads  figure dimensions which you may
change during your work session.

     An improved mechanism to repeatedly typeset troublesome
pieces is now available on the Macintosh; it is called LinoTeX;
see the same ftp sources.  It could be set up on many types
of computer.

     Before using labelfig.tex to attach labels to a graphics
object inserted using boxedeps.tex or BoxedArt.tex, make it a
firm rule to carefully adjust the bounding box using the trimming
commands of these packages, and also at least tentatively scale
and position the object. Beware of changing the grid inadvertently
after the labels have been positioned.  For example, correcting
the bounding box of a PostScript graphics object can foul up the
labels by changing the coordinate grid to which the labels are
attached. This is particularly true for the trimming  commands of
boxedeps.tex and BoxedArt.tex. However, as noted already, change
of scale is much less disruptive, and modest adjustments should be
well tolerated.

     Sometimes the labels protrude so far from the bounding box
of a figure that the figure has to be repositioned.  Best do this
by ad hoc spacing, say using \hglue and \vglue; altering the
bounding box would create a vicious circle.

     Remember that you are responsible for preventing labels
from overlapping. You are responsible for all label typography
including size and style. A label is really just about anything
that can be put in a TeX box. Note that spaces at the beginning
and end of labels will normally be suppressed; if you really want
them you must protect them with TeX braces.

     This package temporarily sets the \mathsurround parameter
of TeX to zero  while the labels are being affixed. This is done
because nonzero \mathsurround space would influence the position
of left and right aligned labels; then, when a texpert or printer
modifies mathsurround, diagram labeling might be disastrously
altered. There is a small price to pay involving labels that are
formatted as caption boxes including mathematics: you  may want or
need to specify an explicit mathsurround space within the caption
box; it will not influence anything outside.

     Those hostile to the use of * as separator between
the X and Y coordinates of label insertion points, are free to
impose another using \SetXYSeparator{<the new separator>}.  
Americans may prefer "," to "*" since they never use a 
comma as a decimal point; on the other hand, * may be more visible.

APPENDIX (I)  MERGING labelfig.tex LABELS INTO AN EPSF GRAPHICS OBJECT.

     As promised in the introduction, here is a recipe useful for
publishers. It works at least on Macintosh and at least for vectorized
graphics and Adobe type1 fonts.  (There is surely a similar recipe for
PCs under MSWindows.)

 (a)  Use boxedeps.tex utility to integrate the figure given by the eps
file, "x.eps" say, with a visible frame around it.  See
\ShowDisplacementBoxes command in boxedeps.tex.  To get precise results
automatically it is important to use the \Trim... commands of
boxedeps.tex making the "DisplacementBox" neatly fit the figure.

 (b)  Use the TeX printer driver and LaserWriter (versions >= 8.1.1) to
export to an EPSF the DVI page containing the integrated, labelled
figure. You now have an EPS file  "xx.eps"  that contains too much, and at
the wrong scale, and at wrong position.

 (c)  Convert the EPSF to an Adode Illustrator format EPSF using
the shareware utility called epsConvert by Sam Weiss
1993-- (currently $25).

 (d)  In Illustrator (or a compatible program), group the labels and the
"DisplacementBox"; copy them to the clipboard and paste them into "x.ps".
This step requires that all the label fonts be "visible to the Macintosh.

 (e)  Translate and scale the pasted group consisting of the labels plus
the "DisplacementBox" so as to make the "DisplacementBox" the bounding
box of (labelless) figure represented by "x.eps".  At this point the
labels will be correctly placed on the figure "x.eps".

 (f)  Ungroup and delete the "DisplacementBox".  The result is the
desired single EPS file, "x+.eps" say, It contains the original figure
plus its labels.  

     Using grouping and ungrouping appropriately in "x+.eps", a
publisher's staff can very efficiently improve label positions etc.

APPENDIX II)  SOME EXOTIC APPLICATIONS

     The grid of labelfig.tex is analogous to a light-table in
classical page makeup with wax or latex glue.  In principle, you
can use it to compose any page from its indivisible parts.  This
even has some of the artisanal charm of classical paste-up
provided you have a fast screen preview to make the process
"interactive".

     In practice labelfig.tex is a tool for nonstandard jobs.
Here are a few going beyond the labelling already discussed.

(I)  GRAPHICS INTEGRATION.

     This is accomplished by treating the imported graphics
objects as labels.  The underlying graphics object is then
typically an empty  \vbox to <dimension>{\vfill} in a TeX
\midinsert...\endinsert construction.  A label line
might be of the form

   (.1*.1) \\

The exact form of the special command varies from driver to
driver.  However, in the case of encapsulated PostScript graphics
(EPSF norm), by relying on boxedeps.tex, one can have the
following standard syntax (independant of driver  (see
boxedeps.doc for details.
  
  (.1*.1) \BoxedEPSF{MyFigure scaled <scale in mils>}\\

This may be slow since it requires TeX to read the PostScript
file to read bounding box using many complex macros.  So you
may want to try

  (.1*.1) \EPSFSpecial{MyFigure}{<scale in mils>}\\

which is fast and driver independant, but it squashes the
bounding box, normally to its lower left corner.

     Similarly for graphics of the Macintosh PICT norm ---
using BoxedArt.tex (same sources) in place of boxedeps.tex.

     This approach to integration is to be recommended when
one is assembling a composite graphics object.

 (II)  COMMUTATIVE DIAGRAM ENHANCEMENT

     Commutative diagrams or arrays of mathematical objects
connected by arrows of various sorts are common in mathematics.
The mathematical objects require the use of TeX.  Recently TeX
acquired a good collection of arrows of all slopes --- that of
LamSTeX --- plus pwerful macros to build the diagrams.

     However, even the LamSTeX collection is often
inadequate; it lacks for example double shafted arrows, dotted
arrows and curved arrows. Fortunately it is possible to produce
such arrows on an individual basis using sophisticated graphics
programs such as Illustrator and AldusFreehand (both serving
the EPSF norm) or using Metafont (with its public domain norm).
Since the creation of each new arrow is a work of love, you
probably want to limit the number of arrows by using LamSTeX
for most arrows. The 40K commutative diagram module of LamSTeX
has been adapted to work with AmSTeX and a copy may be posted
with LabelFig and related files. Unfortunately no one has yet
offered a version that works with Plain TeX or LaTeX.

       Suffice it here to say that when the exotic arrow has
been somehow imported into TeX, labelfig.tex treats it as a
label that one affixes to the commutative diagram.  Two other
steps will be treated in separate notes, namely the matter of
extracting the dimension specifications for the arrow and the
construction of the arrow --- for these steps are far from
unique and often depend intimately on your computer environment. 
Notes for the Macintosh-Textures-Illustrator combination are
found in the file ExoticArrows.doc.

 (III) NESTING 

Ingenuity pays off in exploiting labelfig.tex. One can
mix graphics and typography quite freely.  labelfig.tex is good
for freeform or overlapping arrangements, while boxedeps.tex (or
BoxedArt.tex) is best for regimented non-overlapping
arrangements --- and the two can be combined.

     The default behavior of labelfig.tex is not ideal 
for nesting objects, because to prevent trouble for beginners
the register for labels is globally cleared when \AffixLabels
concludes.  But there are switches available

      \LabelsGlobal      \LabelsLocal

which change this.  To understand this, extend the above test 
by something like:

 \LabelsLocal

 \SetLabels
    (.5*.5) AAA\\
 \endSetLabels

 {
 \SetLabels
    (.5*.5) ZZZ\\
 \endSetLabels
   \AffixLabels{\FirstQuadrant}
 }

   \AffixLabels{\FirstQuadrant}

     There are however potential pitfalls.  Neither
labelfig.tex nor boxedeps.tex has been tested under extreme
conditions. Problems may occur if their procedures are
indiscriminately nested. For boxedeps.tex (not labelfig.tex)
there is a precise cause for worry, namely many of its
variables are "global", which means that TeX braces will not
provide the protection one might expect.

COMMAND SUMMARY FOR labelfig.tex

  Here [...] means optional (one or zero)
       [...]* means any number of such constructs

  \SetLabels
    [[<P>](<X><Sep><Y>) <label> \\]*
  \endSetLabels
  \ShowGrid  
  \AffixLabels{<the figure>}

   --- <P> is tack position, one of eleven or empty
              order irrelevant

                   \L\T      \T      \R\T

                   \L\E      \E      \R\E

                     \L               \R

                   \L\B      \B      \R\B

   --- (<X><Sep><Y>) insertion point;
  <Sep> is separator, = * by default;
  \SetXYSeparator{<Sep>} changes it.
   <X> and <Y> are real numbers

  --- <label> a label to attach 

  --- <the figure> the figure to label 

  \GlobalLabels (default)     
  \LocalLabels  setting for nested constructs.

 \Grids makes ALL grids appear; \HideGrid then makes just next disappear.
 \noGrids returns to default.  The commands are always global.

 \GridLineWidth{<dimension>} adjusts width of grid lines. Default is very
small, to give "hairline" effect. If your grid lines are missing try
setting \GridLineWidth{1pt}.

 \Edges#1 globally changes the edge size of all grids to the numerical 
value #1, which must be 1, 10, 100, or 1000.  The default is 1.

VERSION HISTORY.
 --- Jan 1993: \Edges#1 and [??] option after \SetLabels
 --- July 1992: \Grids, \noGrids, \HideGrid;
       Gridlines become hairlines; \GridLineWidth{<dimension>}.
 --- Oct 1991, Jan 1992: \SetXYSeparator{<Sep>},  \LabelsGlobal,
       \LabelsLocal.
 --- July 1991: first release

Address for bugs and other feedback:

        Raymond S\'eroul
        IREM and Lab. de Typographie Informatise
        Univ. Rene Descartes
        Strasbourg

    Tel 33-88-41-63-45
    Email:  A18645@FRCCSC21.BITNET

        Laurent Siebenmann
        Mathematique, Bat. 425,
        Univ de Paris-Sud,
        91405-Orsay,
        France

    Tel 33-1-6941-7949; 
    Email: lcs@topo.math.u-psud.fr

\def\scalefig#1{\epsfxsize #1\textwidth}

\newtheorem{theorem}{Theorem}
\newtheorem{lemma}{Lemma}
\newtheorem{definition}{Definition}

\newtheorem{corollary}{Corollary}

\newtheorem{algorithm}{Algorithm}

\newcommand {\Rbb}{{\mathbb{R}}}
\newcommand {\Cbb}{{\mathbb{C}}}

\title{{\huge On the Pareto-Optimal
Beam Structure and Design for Multi-User MIMO Interference Channels}}

\author{
Juho Park, {\em Student~Member, IEEE} and Youngchul
Sung$^*$\thanks{$^*$Corresponding author}, {\em
Senior~Member, IEEE}
\thanks{The authors are with the Dept. of Electrical Engineering, KAIST, Daejeon, 305-701, South Korea. E-mail: \{jhp, ycsung\}@kaist.ac.kr.
This research was supported by the Basic Science Research Program through the National Research Foundation of Korea (NRF) funded by the Ministry of Education, Science and Technology (2010-0021269). }
\begin{center}
EDICS: MSP-MULT
\end{center}
}

\begin{document}

\maketitle

\begin{abstract}
In this paper, the Pareto-optimal beam structure for multi-user multiple-input multiple-output (MIMO) interference channels is investigated and
a necessary condition for any Pareto-optimal transmit signal covariance matrix is presented for  the $K$-pair Gaussian $(N,M_1,\cdots,M_K)$ interference channel. It is shown  that any Pareto-optimal transmit signal covariance matrix at a transmitter should have its column space contained in the union of the eigen-spaces of the channel matrices from the transmitter to all receivers. Based on this necessary condition, an efficient parameterization for the beam search space is proposed. The
proposed parameterization is given by the product manifold of a Stiefel manifold and a subset of a hyperplane and enables us to construct a very efficient  beam design algorithm by exploiting its rich geometrical structure and existing tools for optimization on Stiefel manifolds.  Reduction in the beam search space dimension and computational complexity  by the proposed parameterization and the proposed beam design approach is significant when the number of transmit antennas is larger than the sum of the numbers of receive antennas, as in upcoming cellular networks adopting massive MIMO technologies. Numerical results validate the proposed parameterization and the proposed cooperative beam design method based on the parameterization for MIMO interference channels.
\end{abstract}


\section{Introduction}

Multi-user multiple antenna interference channels have gained intensive interest from  research communities in recent years because of the significance  of proper interference control in current and future wireless networks. One of the break-through results in this area is interference alignment by Cadambe and Jafar \cite{Cadambe&Jafar:08IT}, which provides an effective way to achieving maximum degrees-of-freedom (DoF) for MIMO interference channels. However, interference alignment is only DoF optimal, i.e., it is optimal at high signal-to-noise ratio (SNR), whereas in typical cellular networks most receivers experiencing severe interference are located at cell edges and hence operate in the low or intermediate SNR regime. Thus,  Jorswieck {\em
et al.} investigated the multiple antenna interference channel problem from a different perspective based on {\em Pareto-optimality} \cite{Jorswieck&Larsson&Danev:08SP}. The framework of Pareto-optimality is especially useful for interference channels since the users in an interference channel basically form a group for negotiation.  Under this framework,
Jorswieck {\it et al.} showed for multiple-input single-output (MISO) interference channels that any
Pareto-optimal beam vector at a transmitter is a normalized
convex combination of the zero-forcing (ZF) beam vector and  the matched-filtering
(MF)  beam vector in the case of
two users and a linear combination of the channel vectors from the
transmitter to all receivers in the general case of an arbitrary
number of users. Their result and subsequent results by other researchers provide useful parameterizations for the optimal beam search space for efficient cooperative beam design in MISO interference channels \cite{Bjornson&etal:10SP,Zhang&Cui:10SP,Zakhour&Gesbert:10WC,Mochaourab&Jorswieck:11SP,Shang&Chen&Poor:11IT,Park&Lee&Sung&Yukawa:12Arxiv}.  However,  not many results for the Pareto-optimal beam structure for MIMO interference channels are available, although
there exist some results in limited circumstances \cite{Bjornson&etal:12SP,Chen&etal:12TAC, Cao&Jorswieck&Shi:12Arxiv}.

In this paper, we provide a necessary condition for Pareto-optimal beamformers for the {\it $K$-pair Gaussian $(N,M_1,$ $\cdots,M_K)$  interference channel,}\footnote{In the $K$-pair Gaussian $(N,M_1,\cdots,M_K)$  channel, we have $K$ transmitter-receiver pairs, and every transmitter has $N$ transmit antennas and receiver $i$ has $M_i ~(\in \{1,\cdots,N\})$ receive antennas.} which can model general MIMO interference channels, and show that any Pareto-optimal transmit signal covariance matrix at a transmitter should have its column space contained in the union of the eigen-spaces of the channel matrices from the transmitter to all receivers. Based on this, we provide an efficient parameterization for the beam search space not missing Pareto-optimality  whose dimension is independent of the number $N$ of transmit antennas  and is determined only by $(M_1,\cdots,M_K)$, when $N \ge \sum_{i=1}^K M_i$.  The proposed parameterization is given by the product manifold of {\em a Stiefel manifold} and {\em a subset of a hyperplane} and enables us to construct a very efficient cooperative beam design algorithm by exploiting its rich geometrical structure and existing tools for optimization on Stiefel manifolds. Reduction in the beam search space dimension and computational complexity  by the proposed parameterization and the proposed beam design algorithm is significant, when $N >> \sum_{i=1}^K M_i$ as in upcoming cellular systems adopting massive MIMO technologies \cite{Marzetta:10WC, Rusek&etal:12Arxiv}.  Furthermore, the proposed beam design algorithm does not need to fix the number of data streams for transmission beforehand and it finds an (locally) optimal DoF for a given finite SNR. This is  beneficial because the optimal DoF is not known for a finite SNR in most cases.

\noindent \textbf{Notations and Organization}  ~~~In this paper, we will make use of standard notational
conventions. Vectors and matrices are written in boldface with
matrices in capitals. All vectors are column vectors. For a matrix
$\Abf$, $\Abf^H$, $\|\Abf\|$, $\mbox{tr}(\Abf)$, and
$|\Abf|$ indicate the Hermitian transpose, 2-norm, trace, and determinant of $\Abf$, respectively. $\Abf_{ij}$ or $[\Abf]_{ij}$ denotes the element in the $i$-th row and the $j$-th column of  $\Abf$.
$\Cc(\Abf)$ denotes the column space  of
$\Abf$ and   $\Cc^\perp(\Abf)$ denotes the orthogonal complement of $\Cc(\Abf)$.
  $\Pbf_{\Lc}(\vbf)$ denotes the orthogonal projection of a vector $\vbf$ onto a linear subspace $\Lc$.  $\Pibf_{\Abf}=\Abf(\Abf^H\Abf)^{-1}\Abf^H$ represents the
orthogonal projection onto $\Cc(\Abf)$ and
$\Pibf_{\Abf}^\perp=\Ibf - \Pibf_{\Abf}$.   For matrices $\Abf$ and
$\Bbf$, $\Abf \succcurlyeq \Bbf$ means that
$\Abf-\Bbf$ is positive semi-definite. $\Ibf_n$ stands for the identity matrix of size $n$
(the subscript is omitted when unnecessary).  $[\abf_1,\cdots,\abf_L]$ or
$[\abf_i]_{i=1}^L$ denotes the matrix composed of vectors
$\abf_1,\cdots,\abf_L$ and $\mbox{diag}(a_1,\cdots,d_n)$ denotes the diagonal matrix with elements $a_1,\cdots,a_n$. $\xbf\sim\mathcal{CN}(\mubf,\Sigmabf)$ means that
$\xbf$ is circularly-symmetric complex Gaussian-distributed with mean vector $\mubf$ and covariance matrix $\Sigmabf$. ${\mathbb{R}}$,
${\mathbb{R}}_+$, and ${\mathbb{C}}$ denote the sets of real
numbers, non-negative real numbers, and complex numbers,
respectively. $\Rbb^n$ denotes the $n$-dimensional Euclidean space and $\Cbb^n$ denotes the vector space of all complex $n$-tuples. $\Cbb^{n \times p}$ is the set of all $n \times p$ matrices  with complex elements. For a complex number $a$, $\mbox{Re}\{a\}$ denotes the real part of $a$.

The remainder of this paper is organized as follows.
The system model is described  in Section \ref{sec:systemmodel}. In
Section \ref{sec:necessarycond}, a necessary condition and a parameterization for Pareto-optimal transmit beamformers for MIMO interference channels are provided.
 In Section \ref{sec:beam_design}, a beam design algorithm under the obtained parameterization is presented. Numerical results are provided
in Section \ref{sec:numerical},    followed by
conclusions in Section \ref{sec:conclusion}.

\section{System Model}
\label{sec:systemmodel}

In this paper, we consider a Gaussian interference channel with
$K$ transmitter-receiver pairs, where  every transmitter has
$N$ transmit antennas and receiver $i$ has  $M_i$ receive antennas.
We assume that $M_i \ge 1$, $i=1,\cdots,K$, and $N\ge \max\{M_1,\cdots,M_K\}$.  Due to interference from the unwanted transmitters,   the received signal vector $\ybf_i$ at receiver $i$ is given by
\begin{equation} \label{eq:rec_signal}
 \ybf_{i} = \Hbf_{ii}\sbf_{i} + \sum_{j=1, j \neq i}^K
\Hbf_{ij} \sbf_j + \nbf_i,
\end{equation}
where $\Hbf_{ij}$ denotes the $M_i \times N$ channel
matrix from transmitter $j$ to receiver $i$;  $\sbf_j$ is the
$N \times 1$ transmit signal vector at transmitter $j$ generated from  Gaussian  distribution  $\mathcal{CN}(0,\Qbf_j)$; and $\nbf_i$ is the
 additive Gaussian noise vector at receiver $i$ with distribution ${\mathcal{CN}}(0, \Ibf)$. Here, the transmit signal covariance matrix $\Qbf_j ~(={\mathbb{E}}\{\sbf_j \sbf_j^H\})$ at transmitter $j$ is chosen among the feasible set
\begin{equation}
\mathcal{Q}_j :=
\{\Qbf\in\mathbb{C}^{N\times N}
~:~
 \Qbf\succcurlyeq \mathbf{0},
~\text{tr}(\Qbf)\le P_j, ~\text{and}~ 1 \le \text{rank}(\Qbf) \le M_j \},
\end{equation}
where the rank constraint is imposed to guarantee that the number of transmitted data streams is at least one and is less than or equal to the possible maximum $M_j =\min\{M_j,N\}$ for transmitter $j$, $j=1,\cdots,K$.   Note that any value of degree-of-freedom (DoF) from one to the maximum $M_j$ is feasible within the feasible set $\Qc_j$. From here on, we will call the considered MIMO interference channel {\it the $K$-pair Gaussian $(N,M_1,\cdots,M_K)$ MIMO interference channel}.  The considered  $K$-pair Gaussian $(N,M_1,\cdots,M_K)$ MIMO interference channel model  is especially useful for downlink cooperative transmit beamforming in cellular systems. In the cellular downlink case, the transmitters, i.e., basestations can be equipped with many transmit antennas and the number of transmit antennas can be set to be the same in the phase of network design. On the other hand, each receiver, i.e., a mobile station has one or two receive antennas and furthermore the receivers forming a cooperative beamforming group together with the cooperating basestations may not have the same number of antennas. The $K$-pair $(N,M_1,\cdots,M_K)$ MIMO interference channel model  fits  this situation exactly.

Due to the assumption of $N \ge \max \{M_i,i=1,\cdots,,K\}$, the $M_i \times N$ channel matrix $\Hbf_{ij}$ is a fat matrix (i.e., the number of its columns is larger than or equal to that of its rows) and its singular value decomposition (SVD) is given by
\begin{equation}  \label{eq:channelmtxSVD}
\Hbf_{ij} = \Ubf_{ij}[\Sigmabf_{ij},~\mathbf{0}]
[\Vbf_{ij}^\parallel, \Vbf_{ij}^\bot]^H,
\end{equation}
 where $\Ubf_{ij}\in{\mathbb{C}}^{M_i\times M_i}$ is a unitary matrix;
$\Sigmabf_{ij}\in{\mathbb{C}}^{M_i\times M_i}$ is a diagonal matrix
composed of the singular values of $\Hbf_{ij}$;
$\Vbf_{ij}^\parallel\in{\mathbb{C}}^{N\times M_i}$ is a submatrix composed of
orthonormal column vectors that span the eigen-space of $\Hbf_{ij}^H$; and
$\Vbf_{ij}^\bot\in{\mathbb{C}}^{N\times (N-M_i)}$ is a submatrix composed of
orthonormal column vectors that span the zero-forcing space of $\Hbf_{ij}$.
Thus, $\Hbf_{ij}\Vbf_{ij}^\parallel\neq \mathbf{0}$ and $\Hbf_{ij}\Vbf_{ij}^\bot = \mathbf{0}$. From here on, we shall refer to  $\Cc(\Vbf_{ij}^\parallel)$ and $\Cc(\Vbf_{ij}^\perp)$ as the parallel and vertical spaces of $\Hbf_{ij}^H$ (or simply $\Hbf_{ij}$ with slight abuse of notation), respectively. For the purpose of beam design in later sections, we assume that the channel information is known to all the transmitters.

Under the assumption that interference is treated as noise at each receiver, for a given set of
transmit signal covariance matrices $\{\Qbf_1,\cdots,\Qbf_K\}$ and a given set of realized channel
matrices $\{ \Hbf_{ij}, i,j =1,\cdots, K\}$, the rate of the $i$-th transmitter-receiver pair is given by
\begin{align} \label{eq:R_iOneUser}
R_i(\{\Qbf_1,\cdots,\Qbf_K\})=
\log\Big|\Ibf+
\Big(\Ibf+\sum_{j\neq i}\Hbf_{ij}\Qbf_j\Hbf_{ij}^H\Big)^{-1}
\Hbf_{ii}\Qbf_i\Hbf_{ii}^H\Big|
\end{align}
for $i=1,\cdots, K$. Then, for the given set of realized channel matrices, the achievable rate region of the MIMO interference channel with interference treated as noise  is defined as the union of  rate-tuples that can be achieved by all possible combinations of transmit covariance matrices:
\begin{equation} \label{eq:rate_region}
\Rc :=
\bigcup_{\left\{\substack{\Qbf_i:~\Qbf_i\in{\mathcal{Q}_i}, 1\leq i\leq K} \right\}}
(R_1(\{\Qbf_1, \cdots, \Qbf_K),\ \ldots,\ R_K(\Qbf_1, \cdots, \Qbf_K)).
\end{equation}
The outer boundary of the rate region $\Rc$ in the first
quadrant is called the $\textit{Pareto boundary}$ of $\Rc$ and
it consists of rate-tuples for which the rate of any one user cannot
be increased without decreasing the rate of at least one other user.

 In the rest of this paper, we shall investigate the Pareto-optimal transmit beam structure for the $K$-pair Gaussian $(N,M_1,\cdots,M_K)$ MIMO interference channel and develop an efficient beam design algorithm based on the obtained Pareto-optimal beam structure.

\section{A Necessary Condition for Pareto-Optimality for Transmit Beamforming in MIMO Interference Channels}
\label{sec:necessarycond}

In this section, we provide  a necessary condition for Pareto-optimal
 transmit covariance matrices for  the  $K$-pair Gaussian $(N,M_1,\cdots,M_K)$ MIMO interference channel, which reveal the structure of Pareto-optimal transmit beamformers. The necessary condition is given in the following theorem.

\vspace{0.5em}

\begin{theorem} \label{prop:optimal_structure_k}
For the $K$-pair Gaussian $(N, M_1,\cdots,M_K)$ MIMO interference channel in which the channel matrices $\{\Hbf_{ij}\}$ are randomly realized and interference is treated as noise at each receiver, any Pareto-optimal transmit signal covariance matrix $\Qbf_i^\star$ at transmitter $i$ should satisfy
\begin{equation}  \label{eq:ParetoSubspaceCondition}
\Cc(\Qbf_i^\star) \subseteq
\Cc([\Vbf_{1i}^\parallel,\cdots,\Vbf_{Ki}^\parallel])=\Cc([\Hbf_{1i}^H,\cdots,\Hbf_{Ki}^H]) ~~~\mbox{in all cases}
\end{equation}
and
\begin{equation}  \label{eq:ParetoFullPowerCondition}
\text{tr}(\Qbf_i^\star) = P_i ~~~\mbox{in the case that $N \ge \sum_{i=1}^K M_i$}.
\end{equation}
\end{theorem}

\vspace{0.5em}

\textit{Proof:}  First, we consider the case that $N \ge \sum_{i=1}^K M_i$.  Suppose that the matrix $[\Vbf_{1i}^{\parallel},\cdots, \Vbf_{Ki}^{\parallel}]\in\mathbb{C}^{N\times \sum_i M_i}$ has rank $m$
$(< N)$. \footnote{When $m=N$, the condition (\ref{eq:ParetoSubspaceCondition}) is trivially satisfied since the channel matrices are randomly realized and thus $[\Vbf_{1i}^\parallel,\cdots,\Vbf_{Ki}^\parallel]$ spans the whole ${\mathbb{C}}^N$ space.} Then, there exists an orthonormal basis $\{\ubf_l\}_{l=1}^{N-m}$ that spans
$\Cc^\perp([\Vbf_{1i}^{\parallel},\cdots,$ $\Vbf_{Ki}^{\parallel}])$, i.e.,
\begin{equation}
\Cc^\perp([\Vbf_{1i}^{\parallel},\cdots, \Vbf_{Ki}^{\parallel}])
=
\Cc(\{\ubf_l\}_{l=1}^{N-m}).
\end{equation}
Now, suppose that a set of covariance matrices $\{\Qbf_i, i=1,\cdots,K\}$ is Pareto-optimal (i.e., it
achieves a Pareto boundary point of the achievable rate region $\Rc$) and that $\Cc(\Qbf_i)\not\subseteq \Cc([\Vbf_{1i}^{\parallel},\cdots, \Vbf_{Ki}^{\parallel}])$
at transmitter $i$. Then, we can express $\Qbf_i$ as
\begin{equation}
\Qbf_i =
[\Vbf_{1i}^{\parallel},\cdots, \Vbf_{Ki}^{\parallel}]
\Xbf_i
[\Vbf_{1i}^{\parallel},\cdots, \Vbf_{Ki}^{\parallel}]^H
+ \sum_{l=1}^{N-m}\alpha_l^2 \ubf_l\ubf_l^H,
\end{equation}
where $\Xbf_i\succcurlyeq 0$,  $\text{tr}(\Qbf_i)\le P_i$, and $\alpha_l^2 = \ubf_l^H\Qbf_i\ubf_l$. Here, $\Cc(\Qbf_i)\not\subseteq \Cc([\Vbf_{1i}^{\parallel},\cdots, \Vbf_{Ki}^{\parallel}])$ implies that $\alpha_l^2\neq 0$ for some $l \in \{1,\cdots, N-m\}$.  Let $\hat{i}$ be such an index  and  let
\begin{equation}  \label{eq:traceQbfiprime}
\Qbf_i^\prime :=
\Qbf_i - \alpha_{\hat{i}}^2\ubf_{\hat{i}}\ubf_{\hat{i}}^H
\end{equation}
with $\alpha_{\hat{i}}^2\neq 0$. Then,
$\text{tr}(\Qbf_i^\prime) = \text{tr}(\Qbf_i) - \alpha_{\hat{i}}^2 < \text{tr}(\Qbf_i)\le P_i$ and
$\Qbf_i^\prime$ is positive semi-definite.\footnote{The positive semi-definiteness of $\Qbf_i^\prime$ can be shown as in \cite{Lindblom&Larsson&Jorswieck:10WCOM}. First,
$\ubf_{\hat{i}}^H\Qbf_i^\prime\ubf_{\hat{i}} =
\ubf_{\hat{i}}^H(\Qbf_i-\alpha_{\hat{i}}^2\ubf_{\hat{i}}\ubf_{\hat{i}}^H) \ubf_{\hat{i}}
=\ubf_{\hat{i}}^H\Qbf_i\ubf_{\hat{i}} - \alpha_{\hat{i}}^2\|\ubf_{\hat{i}}\|^2
= 0$ by the definition of $\alpha_{\hat{i}}^2$ and $||\ubf_{l}||=1$. For any vector $\wbf$ orthogonal to $\ubf_{\hat{i}}$, we have
$\wbf^H\Qbf_i^\prime\wbf
=
\wbf^H(\Qbf_i-\alpha_{\hat{i}}^2\ubf_{\hat{i}}\ubf_{\hat{i}}^H)\wbf
=
\wbf^H\Qbf_i\wbf \ge 0$ by the positive semi-definiteness of $\Qbf_i$. Since any vector in $\Cbb^N$ is contained in $\Cc([\Vbf_{1i}^{\parallel},\cdots, \Vbf_{Ki}^{\parallel}]) \oplus \Cc(\{\ubf_l\}_{l=1}^{N-m})$. The claim follows.}
Thus, $\Qbf_i^\prime$ is a valid transmit signal covariance matrix. Now consider the rate-tuple that is achieved by   $\{\Qbf_1,\cdots,\Qbf_i^\prime,\cdots,\Qbf_K\}$. Let the interference covariance matrix at receiver $i$ be denoted by
\begin{equation}
\Phibf_i := \Ibf+\sum_{k\neq i}\Hbf_{ik}\Qbf_{k}\Hbf_{ik}^H.
\end{equation}
Then, with the new set of transmit signal covariance matrices, the rate of the $i$-th transmitter-receiver pair  is given by
\begin{align}
R_i(\{\Qbf_1,\cdots,\Qbf_i^\prime,\cdots,\Qbf_K\})
&=
\log \Big|\Ibf+
\Phibf_i^{-1}
\Hbf_{ii}\Qbf_i^\prime\Hbf_{ii}^H \Big| \nonumber \\
&=
\log \Big|\Ibf+
\Phibf_i^{-1}
\Hbf_{ii}(\Qbf_i-\alpha_{\hat{i}}^2\ubf_{\hat{i}}\ubf_{\hat{i}}^H)
\Hbf_{ii}^H \Big| \nonumber\\
&\stackrel{(a)}{=}
\log \Big|\Ibf+
\Phibf_i^{-1}
\Hbf_{ii}\Qbf_i\Hbf_{ii}^H \Big|
\nonumber \\
&=
R_i(\{\Qbf_1,\cdots,\Qbf_i,\cdots,\Qbf_K\}), \label{eq:theo1Rip}
\end{align}
where  step (a) holds because $\ubf_{\hat{i}} \in\Cc^\perp([\Vbf_{1i}^{\parallel},\cdots, \Vbf_{Ki}^{\parallel}])$ and hence $\Hbf_{ii}\ubf_{\hat{i}}={\mathbf{0}}$. Similarly, the rate of the $j$-th transmitter-receiver pair  ($j\neq i$)
is given by
\begin{align}
R_j(\{\Qbf_1,\cdots,\Qbf_i^\prime,\cdots,\Qbf_K\})
&=
\log\Big|\Ibf+
\big(\Ibf+\sum_{\substack{k\neq j, k\neq i}}  \Hbf_{jk}\Qbf_k\Hbf_{jk}^H+\Hbf_{ji}\Qbf_i^\prime\Hbf_{ji}^H\big)^{-1}\Hbf_{jj}\Qbf_j\Hbf_{jj}^H \Big|  \nonumber \\
&=
\log\Big|\Ibf+
\big(\Phibf_j-\alpha_{\hat{i}}^2\Hbf_{ji}\ubf_{\hat{i}}
\ubf_{\hat{i}}^H\Hbf_{ji}^H\big)^{-1}
\Hbf_{jj}\Qbf_j\Hbf_{jj}^H \Big|  \nonumber\\
&\stackrel{(b)}{=}
\log\Big|\Ibf+
\Phibf_j^{-1}\Hbf_{jj}\Qbf_j\Hbf_{jj}^H \Big|  \nonumber \\
&=
R_j(\{\Qbf_1,\cdots,\Qbf_i,\cdots,\Qbf_K\}), \label{eq:theo1Rjp}
\end{align}
where  step $(b)$ holds again because $\ubf_{\hat{i}}\in\Cc^\perp([\Vbf_{1i}^{\parallel},
\cdots, \Vbf_{Ki}^{\parallel}])$ and hence $\Hbf_{ji}\ubf_{\hat{i}}={\mathbf{0}}$. Therefore, the rate-tuple does not change by replacing $\{\Qbf_1, \cdots,\Qbf_i, \cdots, \Qbf_K\}$ with $\{\Qbf_1, \cdots,\Qbf_i^\prime, \cdots, \Qbf_K\}$.

Now, construct another transmit signal covariance matrix $\Qbf_i^{\prime\prime}$ as
\begin{equation}
\Qbf_i^{\prime\prime} := \Qbf_i^\prime + \delta\vbf\vbf^H,
\end{equation}
where $\vbf$ satisfies $\Hbf_{ii}\vbf\neq \mathbf{0}$ while $\Hbf_{ji}\vbf = \mathbf{0}$ for all $j\neq i$.  Such $\vbf$ exists almost surely in $\Cc([\Vbf_{1i}^{\parallel},\cdots, \Vbf_{Ki}^{\parallel}])$ (i.e., $\vbf \in \Cc(\Vbf_{ii}^\parallel) \bigcap \left(\bigcup\limits_{j\neq i} \Cc(\Vbf_{ji}^{\parallel})\right)^\perp  $) for randomly realized channel matrices, because the event
 ${\Cc(\Vbf_{ii}^{\parallel}) \subseteq \bigcup\limits_{j\neq i} \Cc(\Vbf_{ji}^{\parallel})}$ has measure zero.\footnote{The dimension of $\Cc(\Vbf_{ii}^{\parallel})$ is $M_i ~(\ge 1)$  and  the dimension of $\bigcup\limits_{j\neq i} \Cc(\Vbf_{ji}^{\parallel})$ is at most $\sum_{j\ne i} M_j$ which is strictly less than $N$ by the assumption $\sum_i M_i \le N$. The probability that a randomly realized subspace of $\Cbb^N$ is contained in another randomly realized subspace of $\Cbb^N$ with dimension strictly less than $N$ is zero.}
Here, $\delta > 0$ is chosen so that  $\delta \le \frac{1}{\text{tr}(\vbf\vbf^H)}
\Big(P_i-\text{tr}(\Qbf_i^\prime)\Big)$ (this is possible since $\text{tr}(\Qbf_i^\prime) < \text{tr}(\Qbf_i) \le P_i$. See (\ref{eq:traceQbfiprime}).) and
\begin{equation}
    \text{tr}(\Qbf_i^{\prime\prime})
=   \text{tr}(\Qbf_i^{\prime}+\delta\vbf\vbf^H)
\le \text{tr}(\Qbf_i^{\prime})+(P_i - \text{tr}(\Qbf_i^\prime))
= P_i.
\end{equation}
Thus, $\Qbf_i^{\prime\prime}$ is a valid transmit signal covariance matrix.
Now consider the rate-tuple that is achieved by
$\{\Qbf_1,\cdots,$ $\Qbf_i^{\prime\prime},\cdots, \Qbf_K\}$.  Here, we define
\begin{equation}
\Psibf_j :=\Ibf+\sum_{k\neq j, k\neq i }\Hbf_{jk}\Qbf_k\Hbf_{jk}^H
+\Hbf_{ji}\Qbf_i^\prime\Hbf_{ji}^H.
\end{equation}
Then, the rate of the $j$-th transmitter-receiver pair receiver ($j \neq i$) is given by
\begin{align}
R_j(\{\Qbf_1,\cdots,\Qbf_i^{\prime\prime},\cdots, \Qbf_K\})
&=
\log \Big|\Ibf+
\big(\Ibf+\sum_{k\neq i, k\neq j} \Hbf_{jk}\Qbf_k\Hbf_{jk}^H+\Hbf_{ji}\Qbf_i^{\prime\prime}\Hbf_{ji}^H\big)^{-1}
\Hbf_{jj}\Qbf_j\Hbf_{jj}^H\Big| \nonumber\\
&=
\log \Big|\Ibf+
\big(\Psibf_j+\delta\Hbf_{ji}\vbf\vbf^H\Hbf_{ji}^H\big)^{-1}
\Hbf_{jj}\Qbf_j\Hbf_{jj}^H\Big| \nonumber\\
&\stackrel{(c)}{=}
\log \Big|\Ibf+
\Psibf_j^{-1}
\Hbf_{jj}\Qbf_j\Hbf_{jj}^H\Big| \nonumber\\
&= R_j(\{\Qbf_1,\cdots,\Qbf_i^\prime,\cdots,\Qbf_K\}) \nonumber\\
&\stackrel{(d)}{=}R_j(\{\Qbf_1,\cdots,\Qbf_i,\cdots,\Qbf_K\}), \label{eq:theorem1secondRj}
\end{align}
where  step $(c)$ holds by the construction of $\vbf$ and  step (d) holds by (\ref{eq:theo1Rjp}). On the other hand, the rate of the $i$-th transmitter-receiver pair  with  $\{\Qbf_1,\cdots,\Qbf_i^{\prime\prime},$ $\cdots, \Qbf_K\}$ is given by
\begin{align}
R_i(\{\Qbf_1,\cdots,\Qbf_i^{\prime\prime},\cdots,\Qbf_K\})
&=
\log \big|\Ibf+\Phibf_i^{-1}
\Hbf_{ii}\Qbf_i^{\prime\prime}\Hbf_{ii}^H\big| \nonumber\\
&\stackrel{(e)}{=}
\log \big|\Phibf_i+\Hbf_{ii}\Qbf_i^{\prime\prime}\Hbf_{ii}^H\big|
-
\log \big|\Phibf_i\big| \nonumber\\
&=
\log \big|\Phibf_i
+ \Hbf_{ii}(\Qbf_i^\prime+\delta\vbf\vbf^H)\Hbf_{ii}^H\big|
- \log \big|\Phibf_i\big|\nonumber\\
&\stackrel{(f)}{>}
\log \big|\Phibf_i+\Hbf_{ii}\Qbf_i^{\prime}\Hbf_{ii}^H\big|
-
\log \big|\Phibf_i\big|\nonumber\\
&=
\log \Big|\Ibf+\Phibf_i^{-1}
\Hbf_{ii}\Qbf_i^\prime\Hbf_{ii}^H\Big| \nonumber\\
&= R_i(\{\Qbf_1,\cdots,\Qbf_i^\prime,\cdots,\Qbf_K\}) \nonumber\\
&\stackrel{(g)}{=} R_i(\{\Qbf_1,\cdots,\Qbf_i,\cdots,\Qbf_K\}),  \label{eq:theorem1secondRi}
\end{align}
where
 step $(e)$ holds by $|\Ibf+\Abf^{-1}\Bbf|=|\Abf^{-1}||\Abf+\Bbf|$,   step $(f)$ holds by Lemma \ref{lemma:strictEigenvalue}, and  step (g) holds by (\ref{eq:theo1Rip}). This contradicts our assumption that the set $(\Qbf_1,\cdots,\Qbf_i,\cdots,\Qbf_K)$ of transmit signal covariance matrices is Pareto-optimal. Therefore, we have
\[
\Cc(\Qbf_i^\star) \subseteq
\Cc([\Vbf_{1i}^\parallel,\cdots,\Vbf_{Ki}^\parallel]).
\]

Next, suppose that $\Cc(\Qbf_i) \subseteq \Cc([\Vbf_{1i}^\parallel,\cdots,\Vbf_{Ki}^\parallel])$ but $\mbox{tr}(\Qbf_i) < P_i$. Then, by the same argument as before, there almost surely exists
$\vbf$ such that $\Hbf_{ii}\vbf\neq \mathbf{0}$  and $\Hbf_{ji}\vbf = \mathbf{0}$ for all $j\neq i$, when $N \ge \sum_i M_i$. Let
\begin{equation}
\bar{\Qbf}_i = \Qbf_i + \bar{\delta}\vbf\vbf^H,
\end{equation}
where  $\bar{\delta}$ is chosen to be $\bar{\delta} = \frac{1}{\text{tr}(\vbf\vbf^H)}
\Big(P_i-\text{tr}(\Qbf_i)\Big)$ so that $\mbox{tr}(\bar{\Qbf_i})=P_i$. Then, the rate of the $j$-th transmitter-receiver pair ($j \ne i$) does not change by the same argument as in (\ref{eq:theorem1secondRj}) and the rate of the $i$-th transmitter-receiver pair strictly increases by the same argument as in  (\ref{eq:theorem1secondRi}). Thus, in the case of $N \ge \sum_i M_i$, each transmitter should use full power for Pareto optimality.

Now, consider the case of $N  < \sum_{i=1}^K M_i$. In this case, $\Cc([\Vbf_{1i}^\parallel,\cdots,\Vbf_{Ki}^\parallel])= \Cbb^N$ for randomly realized channel matrices and (\ref{eq:ParetoSubspaceCondition}) is trivially true. Finally, $\Cc([\Vbf_{1i}^\parallel,\cdots,\Vbf_{Ki}^\parallel])=\Cc([\Hbf_{1i}^H,\cdots,\Hbf_{Ki}^H])$ by the definition of $\Vbf_{ji}^\parallel$. (See (\ref{eq:channelmtxSVD}).)
$\hfill\blacksquare$

\vspace{0.5em}

\begin{lemma} \label{lemma:strictEigenvalue} Under the same conditions as in Theorem  \ref{prop:optimal_structure_k}, we have
\begin{equation}  \label{eq:lemstrictEigenvalue}
\log \big|\Phibf_i
+ \Hbf_{ii}(\Qbf_i^\prime+\delta\vbf\vbf^H)\Hbf_{ii}^H\big|
> \log \big|\Phibf_i+\Hbf_{ii}\Qbf_i^{\prime}\Hbf_{ii}^H\big|.
\end{equation}

\end{lemma}

\textit{Proof:}  First, consider the difference:
\begin{eqnarray*}
 \left(\Phibf_i
+ \Hbf_{ii}(\Qbf_i^\prime+\delta\vbf\vbf^H)\Hbf_{ii}^H \right) - \left( \Phibf_i+\Hbf_{ii}\Qbf_i^{\prime}\Hbf_{ii}^H \right)
&=& \delta \Hbf_{ii}\vbf\vbf^H\Hbf_{ii}^H  \\
&\succcurlyeq& {\mathbf{0}}.
\end{eqnarray*}
Thus,  $\Phibf_i
+ \Hbf_{ii}(\Qbf_i^\prime+\delta\vbf\vbf^H)\Hbf_{ii}^H \succcurlyeq \Phibf_i+\Hbf_{ii}\Qbf_i^{\prime}\Hbf_{ii}^H$. This implies that the ordered eigenvalues of $\Phibf_i
+ \Hbf_{ii}(\Qbf_i^\prime+\delta\vbf\vbf^H)\Hbf_{ii}^H$ majorize those of $\Phibf_i+\Hbf_{ii}\Qbf_i^{\prime}\Hbf_{ii}^H$. That is, let $\lambda_k^{\prime\prime}$ be the $k$-th largest eigenvalue of $\Phibf_i
+ \Hbf_{ii}(\Qbf_i^\prime+\delta\vbf\vbf^H)\Hbf_{ii}^H$ and let $\lambda_k^\prime$ be the $k$-th largest eigenvalue of $\Phibf_i+\Hbf_{ii}\Qbf_i^{\prime}\Hbf_{ii}^H$. Then,
\begin{equation} \label{eq:lemmaMajorization}
\lambda_k^{\prime\prime} \ge \lambda_k^\prime, ~~~ \forall~k.
\end{equation}
Next, consider the difference of the traces of the two matrices:
\begin{eqnarray}
\mbox{tr}\left(\Phibf_i
+ \Hbf_{ii}(\Qbf_i^\prime+\delta\vbf\vbf^H)\Hbf_{ii}^H \right) - \mbox{tr}\left( \Phibf_i+\Hbf_{ii}\Qbf_i^{\prime}\Hbf_{ii}^H \right)
&=& \delta \mbox{tr}\left(\Hbf_{ii}\vbf\vbf^H\Hbf_{ii}^H \right) \nonumber  \\
&=& \delta || \Hbf_{ii}\vbf ||^2  \nonumber\\
&>& 0 \label{eq:lemma1tracedif}
\end{eqnarray}
by the construction of $\vbf$ satisfying $\Hbf_{ii} \vbf \ne {\mathbf{0}}$.  By (\ref{eq:lemmaMajorization}), (\ref{eq:lemma1tracedif}) and the fact that the trace of a matrix is the sum of its eigenvalues, there exists at least one eigenvalue $\lambda_k^{\prime\prime}$ that is strictly larger than $\lambda_k^\prime$.  Therefore, we have
\[
|\Phibf_i
+ \Hbf_{ii}(\Qbf_i^\prime+\delta\vbf\vbf^H)\Hbf_{ii}^H| >  |\Phibf_i+\Hbf_{ii}\Qbf_i^{\prime}\Hbf_{ii}^H|
\]
since the determinant of a matrix is the product of its eigenvalues and both the matrices are strictly positive-definite due to the added identity matrix in $\Phibf_i$, i.e., $\lambda_k^{\prime\prime} \ge \lambda_k^\prime > 0, ~\forall~k$. Finally, (\ref{eq:lemstrictEigenvalue}) follows by the monotonicity of logarithm.
$\hfill\blacksquare$

\vspace{0.5em}
\noindent Theorem \ref{prop:optimal_structure_k} states that the column space of any Pareto-optimal transmit signal covariance matrix at transmitter $i$ should be contained in the union of the parallel spaces of the channels from transmitter $i$ to all receivers. In the case that $M_i=1$ for all $i=1,\cdots,K$, the parallel space is simply the 1-dimensional linear subspace spanned by the matched filtering vector. Thus, this result in Theorem \ref{prop:optimal_structure_k} can be regarded as a generalization of the result in the MISO interference channel by Jorswieck {\it et al.} \cite{Jorswieck&Larsson&Danev:08SP} to general MIMO interference channels described by the $K$-pair $(N,M_1,\cdots,M_K)$ interference channel model.

\subsection{The Symmetric $2$-User Case}
\label{sec:2-user_case}

In this subsection, we consider the symmetric two-user case and present another
representation for Pareto-optimal transmit signal covariance matrices in this case.

\vspace{0.5em}
\begin{corollary}
In the two-user case in which the number of receive antennas is the same $(M=M_1=M_2)$ and $N\ge 2M$,  any Pareto-optimal transmit signal covariance matrix $\Qbf_1^\star$ at transmitter $1$ should satisfy
\begin{equation}
\Cc(\Qbf_1^\star)
\subseteq
\Cc([\Vbf_{11}^{\parallel}, \Pibf_{\Vbf_{21}^{\bot}} \Vbf_{11}^\parallel])
=
\Cc([\Vbf_{11}^\parallel, \Vbf_{21}^\parallel ])
\end{equation}
and $\mbox{tr}(\Qbf_1^\star)=P_1$, where  $\Pibf_{\Vbf_{21}^{\bot}} \Vbf_{11}^\parallel=(\Vbf_{21}^{\bot}\Vbf_{21}^{\bot H}) \Vbf_{11}^\parallel~\left(=\Pibf_{\Vbf_{21}^\parallel}^{\bot} \Vbf_{11}^\parallel\right)$.
\end{corollary}

\vspace{0.5em}

\textit{Proof:} The proof is by showing the equivalence of the two subspaces:
\begin{equation}\label{eq:equiv_subsp}
\Cc([\Vbf_{11}^{\parallel}, (\Vbf_{21}^\bot\Vbf_{21}^{\bot H}) \Vbf_{11}^\parallel])
=
\Cc([\Vbf_{11}^{\parallel}, \Vbf_{21}^{\parallel}]).
\end{equation}
Any vector in $\Cc([\Vbf_{11}^{\parallel}, \Vbf_{21}^{\parallel}])$ of the right-hand side (RHS) of \eqref{eq:equiv_subsp} can be expressed  as
\begin{equation}
\Vbf_{11}^\parallel \xbf + \Vbf_{21}^\parallel \ybf
\end{equation}
for some $\xbf, \ybf \in \Cbb^M$, whereas  any vector in $\Cc([\Vbf_{11}^{\parallel}, (\Vbf_{21}^\bot\Vbf_{21}^{\bot H}) \Vbf_{11}^\parallel])$ of  the left-hand side (LHS) of \eqref{eq:equiv_subsp} can be expressed as
\begin{equation}\label{eq:rhs_subsp}
\Vbf_{11}^\parallel \xbf^\prime +
(\Vbf_{21}^\bot\Vbf_{21}^{\bot H})\Vbf_{11}^\parallel \ybf^\prime
\end{equation}
for some  $\xbf^\prime, \ybf^\prime \in \Cbb^M$.
Eq. \eqref{eq:rhs_subsp} can be rewritten as
\begin{align}
& \Vbf_{11}^\parallel \xbf^\prime +
 (\Vbf_{21}^\bot\Vbf_{21}^{\bot H})\Vbf_{11}^\parallel \ybf^\prime
 \nonumber \\
=&
 \Vbf_{11}^\parallel \xbf^\prime +
(\Ibf-\Vbf_{21}^\parallel\Vbf_{21}^{\parallel H})\Vbf_{11}^\parallel
 \ybf^\prime
 \nonumber \\
=&
 \Vbf_{11}^\parallel (\xbf^\prime-\ybf^\prime)
-\Vbf_{21}^\parallel(\Vbf_{21}^{\parallel H}\Vbf_{11}^\parallel)
 \ybf^\prime.
\end{align}
Furthermore, $\Vbf_{21}^{\parallel H}\Vbf_{11}^\parallel \in\mathbb{C}^{M\times M}$ is invertible almost surely.\footnote{$\Vbf_{11}^\parallel$ and $\Vbf_{21}^\parallel$ are the parallel spaces of $\Hbf_{11}$ and $\Hbf_{21}$, respectively. The event that $\Vbf_{21}^{\parallel H}\Vbf_{11}^\parallel \in\mathbb{C}^{M\times M}$ is non-invertible requires that $\Cc(\Vbf_{11}^\parallel)$ is contained in a strict subspace of $\Cbb^N$ with dimension less than $N$ determined by $\Vbf_{21}^\parallel$. Such an event has measure zero for randomly realized channel matrices.} Thus, there exists an isomorphism  between $(\xbf,\ybf)$ and $(\xbf^\prime,\ybf^\prime)$ given by
 \begin{equation}
\begin{array}{ll}
\ybf^\prime &= -(\Vbf_{21}^{\parallel H}\Vbf_{11}^\parallel)^{-1}\ybf \\
\xbf^\prime &= \xbf + \ybf^\prime \\
            &= \xbf-(\Vbf_{21}^{\parallel H}
               \Vbf_{11}^\parallel)^{-1}\ybf
\end{array}
\end{equation}
to satisfy
\begin{equation}\label{eq:two_subsp_equal}
 \Vbf_{11}^\parallel \xbf + \Vbf_{21}^\parallel \ybf
=
 \Vbf_{11}^\parallel (\xbf^\prime-\ybf^\prime)
-\Vbf_{21}^\parallel(\Vbf_{21}^{\parallel H}\Vbf_{11}^\parallel)
 \ybf^\prime.
\end{equation}
Thus, the two subspaces are equivalent, i.e., $\Cc([\Vbf_{11}^{\parallel}, \Pibf_{\Vbf_{21}^{\bot}} \Vbf_{11}^\parallel])
=
\Cc([\Vbf_{11}^\parallel, \Vbf_{21}^\parallel ])$. Since $\Cc(\Qbf_1^\star)
\subseteq
\Cc([\Vbf_{11}^\parallel, \Vbf_{21}^\parallel ])$ by Theorem \ref{prop:optimal_structure_k}, the claim follows.
$\hfill\blacksquare$

\vspace{1em} As in the MISO case
\cite{Jorswieck&Larsson&Danev:08SP}, the Pareto-optimal beam space
$\Cc(\Qbf_1^\star)$ is contained in the union of the self-parallel
space of $\Cc(\Vbf_{11}^{\parallel})$ and the vertical or
zero-forcing space $\Cc(\Pibf_{\Vbf_{21}^{\bot}}
\Vbf_{11}^\parallel)$ of the channel to the other user in the
two-user symmetric MIMO case.

\subsection{Parameterization for the Pareto-Optimal Beam Structure in MIMO Interference Channels}

Theorem \ref{prop:optimal_structure_k} provides a necessary condition for Pareto-optimal transmit signal covariance matrices for the $K$-pair Gaussian $(N,M_1,\cdots,M_K)$  MIMO interference channel with interference treated as noise.
Based on Theorem \ref{prop:optimal_structure_k}, in this section, we  develop a concrete parameterization for Pareto-optimal transmit signal covariance matrices for the $K$-pair Gaussian $(N,M_1,\cdots,M_K)$ MIMO interference channel for construction of a very efficient beam design algorithm in the next section. Here, we mainly focus on the case of $N \ge \sum_{i=1}^K M_i$, although the parameterization result here can be applied to the case of $N < \sum_{i=1}^K M_i$.

Since $\Cc(\Qbf_i^\star) \subseteq
\Cc([\Vbf_{1i}^\parallel,\cdots,\Vbf_{Ki}^\parallel])=\Cc([\Hbf_{1i}^H,\cdots,\Hbf_{Ki}^H])$,
any Pareto-optimal transmit signal covariance matrix $\Qbf_i^\star$ at transmitter $i$ can be expressed as
\begin{equation}
\Qbf_i^\star =
[\Hbf_{1i}^H,\cdots,\Hbf_{Ki}^H]
\Xbf_i
[\Hbf_{1i}^H,\cdots,\Hbf_{Ki}^H]^H,
\end{equation}
where  $\Xbf_i$ is a $(\sum_i M_i) \times (\sum_i M_i)$ positive semi-definite matrix with rank less than or equal to $M_i$. Note that $[\Hbf_{1i}^H,\cdots,\Hbf_{Ki}^H]$ is a $N \times (\sum_i M_i)$ matrix and it has full column rank almost surely for randomly realized channels.\footnote{The full column rank assumption is not necessary. In fact, the complexity of the beam design problem is reduced when the matrix does not have full column rank. This step will be explained in Algorithm \ref{algo:overall} in Section \ref{sec:beam_design}.} Let the (skinny) QR factorization of $[\Hbf_{1i}^H,\cdots,\Hbf_{Ki}^H]$ be
\begin{equation}  \label{eq:QRUpsilon}
[\Hbf_{1i}^H,\cdots,\Hbf_{Ki}^H] = \Upsilonbf_i \Rbf_i,
\end{equation}
where $\Upsilonbf_i$ is a $N \times \sum_i M_i$ matrix with orthonormal columns and $\Rbf_i$ is a $(\sum_i M_i) \times (\sum_i M_i)$ upper triangular matrix.  With the QR factorization, the Pareto-optimality subspace condition (\ref{eq:ParetoSubspaceCondition}) can be rewritten as
\begin{equation}  \label{eq:UpXpUpH}
\Qbf_i^\star =
\Upsilonbf_i\Xbf_i^\prime
\Upsilonbf_i^H,
\end{equation}
where $\Xbf_i^\prime$ is a $(\sum_i M_i) \times (\sum_i M_i)$ positive semi-definite matrix with rank less than or equal to $M_i$.  Since $\Xbf_i^\prime$ is Hermitian, i.e., self-adjoint, by the spectral theorem, it has the spectral decomposition given by
\begin{equation}  \label{eq:XipULUH}
\Xbf_i^\prime = \Ubf_i \Lambdabf_i \Ubf_i^H,
\end{equation}
where $\Ubf_i$ is a $(\sum_i M_i) \times M_i$ matrix with orthonormal columns, i.e., $\Ubf_i^H\Ubf_i=\Ibf$ and $\Lambdabf_i= \mbox{diag}(\lambda_{i1},\cdots, \lambda_{iM_i})$ is a $M_i \times M_i$ diagonal matrix with nonnegative elements, i.e., $\lambda_{ik} \ge 0$ for all $k$.  Thus, any Pareto-optimal transmit signal covariance matrix at transmitter $i$ is expressed as
\begin{equation} \label{eq:parameterizationForm}
\Qbf_i^\star = \Upsilonbf_i \Ubf_i \Lambdabf_i \Ubf_i^H \Upsilonbf_i^H,
\end{equation}
which is a spectral decomposition of $\Qbf_i^\star$ since $ (\Upsilonbf_i \Ubf_i)^H (\Upsilonbf_i \Ubf_i)=\Ibf$. Note here that   $\Upsilonbf_i$ is known to the transmitter  under the assumption of known channel information and fixed for a given set of realized channel matrices $\{\Hbf_{ij}\}$.  Note also that  (\ref{eq:parameterizationForm}) incorporates the condition (\ref{eq:ParetoSubspaceCondition}) of Theorem  \ref{prop:optimal_structure_k} only. In the case of $N \ge \sum_i M_i$, we have the full transmission power condition (\ref{eq:ParetoFullPowerCondition}) additionally. Applying this full power constraint to (\ref{eq:parameterizationForm}), we have
\begin{eqnarray}
P_i &=& \mbox{tr}(\Qbf_i^\star) \nonumber\\
&=& \mbox{tr}( \Upsilonbf_i \Ubf_i \Lambdabf_i \Ubf_i^H \Upsilonbf_i^H)= \mbox{tr}(  \Lambdabf_i \Ubf_i^H \Upsilonbf_i^H\Upsilonbf_i\Ubf_i ), ~~~~~(\Upsilonbf_i \Ubf_i)^H (\Upsilonbf_i \Ubf_i)=\Ibf\nonumber\\
&=&  \mbox{tr}( \Lambdabf_i) = \sum_{k=1}^{M_i} \lambda_{ik},  \label{eq:ParamPowerHyper}
\end{eqnarray}
where $\lambda_{ik} \ge 0$ for all $k$.  Thus, any Pareto-optimal transmit signal covariance matrix can be parameterized by the two matrices $\Ubf_i$ and $\Lambdabf_i$ with  constraints $\Ubf_i^H\Ubf_i=\Ibf$ and $\mbox{tr}(\Lambdabf_i)=P_i$, respectively. Especially,  $\Ubf_i$'s satisfying $\Ubf_i^H\Ubf_i=\Ibf$  form a special subset of   $\Cbb^{(\sum_i M_i) \times M_i}$ called the {\it Stiefel manifold $V_{\sum_iM_i,M_i}$}.

\vspace{0.5em}

\begin{definition}[Stiefel manifold] The (compact) Stiefel manifold $V_{n,p}$ (or $V_p(\Cbb^n)$) is the set of all $n \times p$ complex  matrices with orthonormal columns, i.e.,
\begin{equation}
V_{n,p} := \{ \Ubf \in \Cbb^{n\times p}: \Ubf^H \Ubf = \Ibf_p\}.
\end{equation}
\end{definition}

\vspace{0.5em} \noindent Note that $\Cbb^{n\times p}$ is a vector space over $\Cbb$ with the normal matrix addition and the scalar multiplication as vector addition and scalar multiplication. The Stiefel manifold $V_{n,p}$ is a submanifold of the vector space $\Cbb^{n\times p}$ \cite{Absil&etal:book}. Now, we present our parameterization result for Pareto-optimal beamforming in the $K$-pair $(N,M_1,\cdots,M_K)$ MIMO interference channel when $N \ge \sum_i M_i$ in the following theorem.

\vspace{0.5em}

\begin{theorem} \label{theo:parameterization} Any Pareto-optimal transmit signal covariance matrix at transmitter $i$ for the $K$-pair Gaussian $(N,M_1,\cdots,M_K)$ MIMO interference channel with $N \ge \sum_i M_i$ is completely parameterized by the product manifold $\Mc_i$:
\begin{equation}  \label{eq:McIdef}
\Mc_i := V_{\sum_i M_i, M_i} \times \Hc_{M_i},
\end{equation}
where $V_{\sum_iM_i, M_i}$ is the Stiefel manifold of orthonormal $M_i$-frames in $\Cbb^{\sum_i M_i}$ and $\Hc_{M_i}$ is a subset in the first quadrant of a hyperplane in the Euclidean space $\Rbb^{M_i}$ defined by
\begin{equation}
\Hc_{M_i} := \{ (\lambda_1, \cdots, \lambda_{M_i}): \lambda_i \ge 0 ~\mbox{and}~ \sum_i \lambda_i = P_i  \}.
\end{equation}
\end{theorem}

{\it Proof:} Combining Theorem \ref{prop:optimal_structure_k} and equations (\ref{eq:UpXpUpH}), (\ref{eq:XipULUH}), (\ref{eq:parameterizationForm}) and (\ref{eq:ParamPowerHyper}), we have the result. $\hfill\blacksquare$

\vspace{0.5em} \noindent Note that $\Mc_i$ is an embedded manifold within the original high dimensional space $\Qc_i$. The main advantage of the parameterization in Theorem \ref{theo:parameterization} is that {\it the dimension of the parameter space (or beam search space) not losing Pareto-optimality does not depend on the number $N$ of transmit antennas when $N \ge \sum_i M_i$} and the proposed parameterization significantly reduces the dimension of the beam search space as compared to the original search space $\Qc_i$, when $N >> \sum_i M_i$.  Thus,  the proposed parameterization  is useful for upcoming cellular downlink cooperative transmission with massive MIMO technologies \cite{Marzetta:10WC, Rusek&etal:12Arxiv} in which   large-scale transmit antenna arrays  are adopted at basestations while each mobile station still has  a limited number of receive antennas.  The exact dimension of the parameter space $\Mc_i$ for transmitter $i$  is given  by
\begin{equation} \label{eq:paramspaceDimension}
D(\Mc_i) =  2 (\sum_{i=1}^K M_i)M_i - (M_i)^2 + (M_i-1).
\end{equation}
This is because the dimension of $V_{n,p}$ is given by $2np - p^2$ \cite{Absil&etal:book} and the dimension of $\Hc_n$ in $\Rbb^n$ is given by $n-1$.
In addition to the independence of the parameter space dimension on the number of transmit antennas,  the parameterization in Theorem \ref{theo:parameterization} enables us to exploit the rich geometrical structure of Stiefel manifolds and hyperplanes for optimal search for beam design. This will become clear shortly in the next section.

 Now, consider the case that $N < \sum_i M_i$. In this case, Theorem \ref{prop:optimal_structure_k} is not so helpful, but a parameterization similar to  that in Theorem \ref{theo:parameterization} can be obtained by directly applying spectral decomposition to $\Qbf_i^\star \in \Cbb^{N\times N}$ with rank less than or equal to $M_i ~(\le N)$. The spectral decomposition of $\Qbf_i^\star$ in this case is given by
\begin{equation}
\Qbf_i^\star = \Ubf_i \Lambdabf_i \Ubf_i^H,
\end{equation}
where $\Ubf_i$ is a $N \times M_i$ matrix with orthonormal columns, i.e., $\Ubf_i^H\Ubf_i=\Ibf_{M_i}$ and $\Lambdabf_i$ is a $M_i \times M_i$ positive semi-definite diagonal matrix. Thus, the parameter space is given by $\Mc_i^\prime := V_{N,M_i} \times \Hc\Sc_{M_i}$, where $\Hc\Sc_{M_i}$ is a subset of a half space of $\Rbb^{M_i}$, defined as $\Hc\Sc_{M_i}:= \{ (\lambda_1, \cdots, \lambda_{M_i}): \lambda_i \ge 0 ~\mbox{and}~ \sum_i \lambda_i \le P_i  \}$.

\section{The Proposed Beam Design Algorithm}
\label{sec:beam_design}

   In this section, we  provide an efficient beam design algorithm
 under the parameterization $\Mc_i$ containing all Pareto-optimal beamformers in the previous section by exploiting the geometric structure of the parameter space. Here, we consider a centralized beam design approach under the assumption that all channel information is available.  For example, in cellular systems, all channel information from cooperating basestations can be delivered to the basestation combiner (BSC), and the BSC can compute beamforming matrices for all the basestations under its control and inform the computed beamforming matrices to the basestations under its control. When fast communication between the BSC and the basestations is available, such a method can be used in practice.

\subsection{The Overall Algorithm Structure: A Utility Function-Based Approach}

Our approach to beam design is based on the utility function based method in \cite{Jorswieck&Larsson:08ICASSP,Park&Lee&Sung&Yukawa:12Arxiv}. In this approach, we define a  utility function $u$:
\begin{equation}
u: \Rbb_+^K \mapsto \Rbb: (R_1,\cdots,R_K) \mapsto u.
\end{equation}
The utility function is chosen to represent the desired system performance metric. We assume that $u$ is a bounded smooth function of $(R_1,\cdots,R_K)$.  In addition, due to Theorem \ref{theo:parameterization}, we have the following mapping:
\begin{equation}
\rbf: \Mc_1 \times \cdots \times \Mc_K \mapsto (R_1,\cdots,R_K),
\end{equation}
which is determined by the rate formula (\ref{eq:R_iOneUser}) and $\Qbf_i = \Upsilonbf_i \Ubf_i \Lambdabf_i \Ubf_i^H \Upsilonbf_i^H$.
Here, we only need to consider $\Mc_1\times \cdots \Mc_K$ as our beam search space owing to Theorem \ref{theo:parameterization}. The composition of the two mappings is given by
\begin{equation}
\tilde{u}:= u \circ \rbf: \Mc_1 \times \cdots \times \Mc_K \mapsto \Rbb.
\end{equation}
Note that this mapping is the desired mapping from the beam search space containing all Pareto-optimal beams to the set of utility values and that $\tilde{u}$ is a smooth function on the product manifold $\Mc_1 \times \cdots \times \Mc_K$ by the smoothness assumption on $u$ and the smoothness of the rates as functions of $\{\Qbf_i\}$.  Then, the utility-maximizing beam design problem is formulated as
\begin{equation}  \label{eq:untilmaxopt}
\max\limits_{\{(\Ubf_i,\Lambda_i) \in \Mc_i, \forall i\}}\  ~~  \tilde{u} \big(\Ubf_1,\Lambdabf_1,\cdots,\Ubf_K,\Lambdabf_K \big),
\end{equation}
where $\Mc_i$ is given by (\ref{eq:McIdef}).  Although simultaneous optimization of $(\Ubf_1,\Lambdabf_1,\cdots,\Ubf_K,\Lambdabf_K)$ to maximize the utility function is difficult,  the optimization (\ref{eq:untilmaxopt}) can efficiently be solved
by an alternating optimization technique.  That is, we fix all other
$\{(\Ubf_j,\Lambdabf_j), j\ne i\}$ except $(\Ubf_i,\Lambdabf_i)$ and
update the unfixed parameters $(\Ubf_i,\Lambdabf_i)$ in order that the utility function is
maximized.  After this update, the next $(\Ubf_i,\Lambdabf_i)$ is picked for update.  This procedure continues until it converges. The
proposed overall algorithm is described below.

\vspace{1em}
\begin{algorithm} \label{algo:overall} The Proposed Beam Design Algorithm - The Overall Structure

\noindent \textbf{Requirements:}

\begin{itemize}
\item Channel information $\{\Hbf_{ij},i,j=1,\cdots,K\}$

\item Maximum available transmit power $\{P_1,\cdots,P_K\}$

\item Utility function $u(R_1,\cdots,R_K)$

\item Stopping tolerance $\epsilon >0$

\end{itemize}

\noindent \textbf{Preprocessing:}

\begin{itemize}
\item Obtain $\Upsilonbf_i$ by  QR factorization of $[\Hbf^H_{1i},\cdots,\Hbf^H_{Ki}]=:\Hbf_i$ as in (\ref{eq:QRUpsilon}) for all $i=1,\cdots,K$.

\item In the above QR factorization step, the rank of $\Hbf_i$ is revealed. Based on the revealed rank\footnote{If $\Hbf_i$ is not of full column rank,  the problem size simply reduces.} $m_i$, set the number of rows of $\Ubf_i$ as $m_i$ and set the number of its columns as $M_i$. In this step, the proper  Stiefel manifold for $\Ubf_i$ is identified and it is $V_{m_i,M_i}$.

\end{itemize}

\noindent \textbf{Iteration:}

\textbf{Initialization:}

\begin{itemize}
\item $l=0$

\item $\Ubf_i^{(0)} = \left[ \begin{array}{c}  \Ibf_{M_i} \\ {\mathbf{0}}    \end{array}  \right]$  and  $\Lambdabf_i^{(0)} = \frac{P_i}{M_i} \Ibf_{M_i}$ for all $i=1,\cdots,K$

\end{itemize}

\vspace{0.5em}
\textbf{while} $\left|\tilde{u}\left(\{(\Ubf_i^{(l)},\Lambdabf_i^{(l)}\} \right) -\tilde{u}\left(\{(\Ubf_i^{(l-1)},\Lambdabf_i^{(l-1)}\} \right)\right| > \epsilon $

\vspace{0.5em}

    \begin{itemize}
        \item[] $l = l+1;$

        \item[] \textbf{for}~ $i=1,\cdots, K$
        \begin{equation}  \label{eq:untilmaxoptperuser}
         (\Ubf_i^{(l+1)},\Lambdabf_i^{(l+1)})=\arg\max_{(\Ubf_i,\Lambda_i) \in \Mc_i}  ~~ \left.\tilde{u}
          \big(\Ubf_1,\Lambdabf_1,\cdots,\Ubf_K,\Lambdabf_K \big)\right|_{\{(\Ubf_j,\Lambda_j)=(\Ubf_j^{(l)},\Lambda_j^{(l)}),j \ne i\}}
        \end{equation}

        \item[] \textbf{end for}
   \end{itemize}

 \textbf{end while}

\noindent \textbf{Postprocessing:}

\begin{itemize}
\item Check the rank of $\Lambdabf_i^{(l_{stop})}$ to determine the number $d_i$ of data streams for transmitter $i$.

\item Construct a beamformer matrix $\Gammabf_i$ for transmitter $i$ as
\begin{equation} \label{eq:beamformGammai}
\Gammabf_i = \Upsilonbf_i \Ubf_i^{(l_{stop})}(:,1:d_i)\sqrt{\mbox{diag}(\lambda_{i1}^{(l_{stop})},\cdots, \lambda_{id_i}^{(l_{stop})})},
\end{equation}
where $d_i= \mbox{rank}\left(\Lambdabf_i^{(l_{stop})}\right)$ and $\Ubf_i^{(l_{stop})}(:,1:d_i)$ is the matrix composed of the first $d_i$ columns of $\Ubf_i^{(l_{stop})}$.

\item At transmitter $i$, generate $d_i$ zero-mean i.i.d. data streams and construct the $d_i \times 1$ data  vector $\dbf_i$ with the generated data streams. Then, construct the signal vector $\sbf_i = \Gammabf_i \dbf_i$ and transmit through antennas. Then, the signal vector $\sbf_i$ has the desired signal covariance matrix $\Qbf_i$.  Typically, $d_i$ i.i.d. data streams are from $d_i$ independent channel encoders.
\end{itemize}

\end{algorithm}

\vspace{0.5em} \noindent There are several interesting features about the proposed beam design algorithm.

\begin{itemize}

\item First, it is not necessary to predetermine the number of
data streams for the algorithm. Although there exist some
asymptotic results on optimal DoF at high SNR
\cite{Cadambe&Jafar:08IT}, the optimal number of independent data
streams for transmission  is not known for finite SNR in most
cases except the known fact that the maximum DoF for transmitter
$i$ is $M_i$. Our parameterization for the beam search space
includes all possible DoF values less than or equal to $M_i$.
Thus, if the algorithm works properly, the algorithm will find the
optimal DoF for given SNR automatically. When the full DoF  $M_i$
is not optimal, the algorithm would return
$(\lambda_{i1},\cdots,\lambda_{iM_i})$ on a corner or an edge of
$\Hc_{M_i}$.

\item Any transmit signal covariance matrix $\Qbf_i$ can be implemented by a beamforming matrix $\Gammabf_i$ as in (\ref{eq:beamformGammai}).

\item Due to the non-convexity of utility functions with respect to (w.r.t.) $\{\Qbf_i\}$ (note the rate formula (\ref{eq:R_iOneUser})), the convergence of the proposed algorithm to the global optimum is not guaranteed, but the proposed algorithm
converges to a locally optimal point by the monotone convergence
theorem, if the step (\ref{eq:untilmaxoptperuser}) works properly, i.e., at each iteration it finds a better point in $\Mc_i$ than the current point. This is because  the utility function is upper bounded and the
proposed algorithm yields a monotonically increasing sequence of
utility function values under the assumption of proper operation of the step (\ref{eq:untilmaxoptperuser}). Furthermore, in this case the proposed algorithm is stable since it monotonically converges.

\end{itemize}

\noindent Thus, an efficient and successful implementation of the step (\ref{eq:untilmaxoptperuser}) is critical to the overall beam design algorithm. Such an implementation is possible and available because of the geometry of our parameterization $\Mc_i$. The problem (\ref{eq:untilmaxoptperuser}) involves optimization on a Stiefel manifold, which is well established \cite{Edelman:98SIAM, Absil&etal:book}. In the next subsections, we briefly introduce some basic facts about Stiefel manifolds and then present our algorithm  implementing (\ref{eq:untilmaxoptperuser}) based on the steepest descent method or the Newton method on Stiefel manifolds of Edelman {\it et al.} \cite{Edelman:98SIAM}.

\subsection{Preliminaries: Riemannian Geometry on Stiefel Manifolds}

Since  geometry of hyperplanes or half spaces is simple, we here provide some basic facts about the Stiefel manifold $V_{n,p}$ that are necessary to understand the  subalgorithm implementing  the step (\ref{eq:untilmaxoptperuser}). For a detailed explanation of the Stiefel manifold and its  geometry, please refer to \cite{Edelman:98SIAM, Absil&etal:book}.  For general Riemannian geometry,   please refer to \cite{Boothby:book, DoCarmo:book}.

\vspace{0.5em}

\noindent {\it Tangent spaces:} The tangent space $T_\Ubf V_{n,p}$ at a point $\Ubf \in V_{n,p}$ is given by
\begin{eqnarray}
T_\Ubf V_{n,p} &=& \{  \Deltabf: \Deltabf^H\Ubf + \Ubf^H \Deltabf = {\mathbf{0}}\}\\
&=& \{ \Ubf \Abf + \Ubf_\perp \Bbf: \Abf^H = -\Abf, ~\Bbf \in \Cbb^{(n-p)\times p}\},
\end{eqnarray}
where $\Ubf\Ubf^H + \Ubf_\perp \Ubf_\perp^H = \Ibf_n$. That is, a tangent vector at $\Ubf$ is a $n\times p$ matrix $\Deltabf$ s.t. $\Ubf^H\Deltabf$ is skew-Hermitian.

\vspace{0.5em} \noindent {\it The canonical metric:} For two tangent vectors $\Deltabf_1$ and $\Deltabf_2$ in $T_\Ubf V_{n,p}$, the canonical metric is defined as
\begin{equation}
g_c(\Deltabf_1,\Deltabf_2) = \mbox{Re}\left\{\mbox{tr}\left( \Deltabf_1^H ( \Ibf - \frac{1}{2} \Ubf\Ubf^H) \Deltabf_2 \right)\right\}
\end{equation}

\vspace{0.5em} \noindent {\it Geodesics:}  A geodesic on a manifold is a curve on the manifold whose velocity vector field is constant along the curve w.r.t. a given affine connection. A geodesic formula for the Stiefel manifold $V_{n,p}$ w.r.t. the Levi-Civita connection is given by the following theorem by Edelman {\it et al.} \cite{Edelman:98SIAM}:

\begin{theorem}[Edelman {\it et al.}\cite{Edelman:98SIAM}] \label{theo:geodesic} Let $\Ubf$ be a point in $V_{n,p}$  and $\Deltabf$ be a tangent vector in $T_\Ubf V_{n,p}$. Then, the geodesic on the Stiefel manifold emanating from $\Ubf$ in the direction $\Deltabf$ is given by the curve
\begin{equation}  \label{eq:EdelmanGeodesic}
\Ubf(t) = \Ubf \Mbf(t) + \Qbf \Nbf(t),
\end{equation}
where
\begin{equation}
\Qbf\Rbf = (\Ibf - \Ubf\Ubf^H) \Deltabf
\end{equation}
is the skinny QR decomposition of $(\Ibf - \Ubf\Ubf^H) \Deltabf$  with  $\Qbf$ being $n\times p$ and $\Rbf$ being $p \times p$,  and $\Mbf(t)$ and $\Nbf(t)$ are $p\times p$ matrices given by the following matrix exponential
\begin{equation}  \label{eq:matrixExponential}
\left[ \begin{array}{c} \Mbf(t) \\ \Nbf(t) \end{array}  \right] = \exp \left( t \left[ \begin{array}{cc} \Abf & - \Rbf^H \\ \Rbf & {\mathbf{0}} \end{array}  \right]\right) \left[ \begin{array}{c} \Ibf_p\\ {\mathbf{0}} \end{array}  \right],
\end{equation}
where $\Abf = \Ubf^H \Deltabf$.
\end{theorem}

\vspace{0.5em} \noindent {\it Gradient:} For a smooth function $f$ on the Stiefel manifold, i.e., $f:V_{n,p}\rightarrow \Rbb$, the gradient of $f$ at $\Ubf$ w.r.t. the canonical metric is defined as the tangent vector $\mbox{grad} f \in T_\Ubf V_{n,p}$ satisfying
$\mbox{Re}\left\{\mbox{tr}(f_\Ubf^H \Deltabf)\right\} = g_c(\mbox{grad} f,\Deltabf)$ for all tangent vectors $\Deltabf$ at $\Ubf$, where $f_\Ubf$ is the $n\times p$ matrix composed of partial derivatives of $f$ w.r.t. the elements of $\Ubf$, i.e.,
$[f_\Ubf]_{ij} = \frac{\partial f}{\partial \Ubf_{ij}}$. The gradient of $f$ at $\Ubf$ is given by
\begin{equation}  \label{eq:StiefelGradient}
\mbox{grad} f = f_\Ubf - \Ubf f_\Ubf^H \Ubf.
\end{equation}

\vspace{0.5em} \noindent {\it Hessian:} For a general Riemannian manifold $(\Sc,g)$, the Hessian operator of a smooth function $f$ at a point $q \in \Sc$ is defined as a linear operator: $\mbox{Hess}_f:T_q\Sc \rightarrow T_q\Sc$ with $\mbox{Hess}_f(\vbf) = \nabla_\vbf \mbox{grad} f$ for all $\vbf \in T_q\Sc$, where  $\nabla$ is the Levi-Civita connection on $\Sc$. Just as in the Euclidean case, a smooth function on $\Sc$ admits Taylor expansion \cite{Absil&etal:book}. Let $\hat{f}_q := f \circ R_q$, where $R_q$ is a retraction.\footnote{A retraction $R_q$ is a smooth mapping from $T_q\Sc$ to $\Sc$ with $R_q({\mathbf{0}})= q$ and $dR_q|_{\mathbf{0}}$ is an identity map, where $dR_q$ is the differential of $R_q$. The exponential map $\exp_q$ is an example of retraction.} Then, in a neighborhood of $q$, we have
\begin{equation}  \label{eq:quadApprox}
\hat{f}(\vbf) \approx f(q) + g(\mbox{grad}f,\vbf) + \frac{1}{2}g(\mbox{Hess}_f(\vbf),\vbf).
\end{equation}
Thus, the stationary point $\vbf^*$ of the RHS of (\ref{eq:quadApprox}) satisfies the Newton equation:
\begin{equation}  \label{eq:NewtonEq}
\mbox{Hess}_f(\vbf^*) + \mbox{grad}f =0.
\end{equation}
The Hessian operator can be computed for complex Stiefel manifolds as well as real Stiefel manifolds. For detail, please refer to  \cite{Edelman:98SIAM} and \cite{Manton:02SP}.

\subsection{The Subalgorithm: Steepest Descent or  Newton Method on the Product Manifold}

Notice that  the cost function $\tilde{u}$ in \eqref{eq:untilmaxoptperuser} is a smooth mapping from $V_{\sum_iM_i,M_i} \times \Hc_{M_i}$ to $\Rbb$ when $\{(\Ubf_j,\Lambdabf_j), ~\forall~j\ne i\}$ are fixed. By exploiting the product structure of the parameter space, the optimization problem \eqref{eq:untilmaxoptperuser}  can be solved by an alternating technique again. That is, first we fix $\Lambdabf_i$ and update $\Ubf_i$ by the steepest descent or Newton method on the Stiefel manifold $V_{\sum_iM_i,M_i}$ \cite{Edelman:98SIAM, Absil&etal:book}. Next, we fix $\Ubf_i$ and update $\Lambdabf_i$ by the steepest descent or Newton method on $\Hc_{M_i}$. We continue this iteration until we have satisfactory convergence. The subalgorithm implementing  the step \eqref{eq:untilmaxoptperuser} is given below.

\vspace{0.5em}
\begin{algorithm} \label{algo:newton} The Subalgorithm for \eqref{eq:untilmaxoptperuser}

\textbf{Requirements:}

\begin{itemize}
\item Cost function $\tilde{u}(\Ubf_i,\Lambdabf_i)$. Set $f=-\tilde{u}$.

\item Step sizes $\tau_1$ and $\tau_2$

\item Stopping tolerance $\epsilon^\prime$

\end{itemize}

\textbf{Initialization:}

\begin{itemize}
\item $k=0$

\item $(\Ubf_{i,(0)},\Lambdabf_{i,(0)})=(\Ubf_{i}^{(l)},\Lambdabf_{i}^{(l)})$

\end{itemize}

\textbf{while} $\left|\tilde{u}\left(\Ubf_{i,(k)},\Lambdabf_{i,(k)} \right) -\tilde{u}\left(\Ubf_{i,(k-1)},\Lambdabf_{i^(k-1)} \right)\right| > \epsilon^\prime $

\vspace{0.5em}

    \begin{itemize}

        \item[] $k = k+1;$

        \item[] \underline{$U$ step}

        \begin{itemize}

        \item[] Fix $\Lambdabf_i$. Given the current $\Ubf_{i,(k)} \in V_{\sum_iM_i,M_i}$,

        \item[1.]  Compute the movement direction vector $\Dbf \in T_{\Ubf_{i,(k)}} V_{\sum_iM_i,M_i}$.

         \begin{itemize}

         \item For the steepest descent method, $\Dbf:=-\mbox{grad} f$ in (\ref{eq:StiefelGradient}).

         \item For the Newton method, compute $\Dbf$ as in \cite{Edelman:98SIAM,Manton:02SP}.

         \end{itemize}

        \item[2.]  Move from $\Ubf_{i,(k)}$ to $\exp_{\Ubf_{i,(k)}}(\tau_1\Dbf)$, where $\exp_{\Ubf_{i,(k)}}(\cdot)$ is the exponential map at $\Ubf_{i,(k)}$. That is, move from  $\Ubf_{i,(k)}$ in direction $\Dbf$ to $\Ubf(\tau_1)$ in (\ref{eq:EdelmanGeodesic}) along the geodesic given by Theorem \ref{theo:geodesic}. Then,
          $\Ubf_{i,(k+1)}= \Ubf(\tau_1)$.

   \end{itemize}

        \item[] \underline{$\Lambda$ step}

         \begin{itemize}

        \item[]    Fix $\Ubf_i$. Given the current $\Lambdabf_{i,(k)} \in \Hc_{M_i}$,

         \item[1.]  Compute the movement direction vector $\etabf$.

         \begin{itemize}

         \item For the steepest descent method, compute the gradient vector $\gbf$ of $f(\Lambdabf_i)$ at $\Lambdabf_{i,(k)}$, and $\etabf := -\gbf$.

         \item For the Newton method, compute the Hessian matrix $\Hbf$ of $f(\Lambdabf_i)$ at $\Lambdabf_{i,(k)}$, and $\etabf:=-\Hbf^{-1}\gbf$

         \end{itemize}

        \item[2.]  Obtain the projection $\Pbf_{T_{\Lambda_{i,(k)}}\Hc_{M_i}}(\etabf)$ of  $\etabf$ to the tangent space $T_{\Lambda_{i,(k)}}\Hc_{M_i}$.

        \item[3.]  Move from $\Lambdabf_{i,(k)}$ to the direction $\etabf$ on  $\Hc_{M_i}$. That is,
        \begin{equation}  \label{eq:lambdaupdate}
           \Lambdabf_{i,(k+1)}= \Pbf_{\Hc_{M_i}}[\Lambdabf_{i,(k)} + \tau_2\Pbf_{T_{\Lambda_{i,(k)}}\Hc_{M_i}}(\etabf)].
        \end{equation}

   \end{itemize}

   \end{itemize}

 \textbf{end while}

\begin{itemize}
\item $(\Ubf_{i}^{(l+1)},\Lambdabf_{i}^{(l+1)})=(\Ubf_{i,(k_{stop})},\Lambdabf_{i,(k_{stop})})$

\end{itemize}

\end{algorithm}

\vspace{0.5em}
\noindent The step 3 in the $U$ step is to maximize the utility with the constraint that the points still stay in the Stiefel manifold $V_{\sum_iM_i,M_i}$. Note that for the $\Lambda$ step, $\tilde{u}(\Lambdabf_i)$ is a conventional multi-variable scalar function, i.e., it is  $\tilde{u}(\lambda_{i1},\cdots,\lambda_{iM_i})$. Thus, the ordinary gradient vector and the ordinary Hessian matrix for a function defined on a Euclidean space are valid. Furthermore, the $\Lambda$ step is simple since a hyperplane is flat and thus its geometry is induced by projection from its embedding Euclidean space. In (\ref{eq:lambdaupdate}), $\Lambdabf_{i,(k)} + \Pbf_{T_{\Lambda_{i,(k)}}\Hc_{M_i}}(\Hbf^{-1}\gbf)$ is still on the hyperplane containing $\Hc_{M_i}$ but it may be outside $\Hc_{M_i}$ (i.e., not in the first quadrant). Projection back to $\Hc_{M_i}$ can be done by simple scaling of $\Pbf_{T_{\Lambda_{i,(k)}}\Hc_{M_i}}(\Hbf^{-1}\gbf)$ after checking the coordinate values of  $\Lambdabf_{i,(k)} + \Pbf_{T_{\Lambda_{i,(k)}}\Hc_{M_i}}(\Hbf^{-1}\gbf)$. That is, if there exists a negative value at some coordinate, $\Pbf_{T_{\Lambda_{i,(k)}}\Hc_{M_i}}(\Hbf^{-1}\gbf)$ is scaled down and then added to $\Lambdabf_{i,(k)}$ so that the value at that coordinate becomes zero.

An attracting aspect of the steepest descent method and the Newton method on the Stiefel manifolds is that their local convergence is established \cite{Absil&etal:book}. Thus, Algorithm \ref{algo:newton} has the local convergence property and therefore, the overall algorithm, Algorithm \ref{algo:overall}, has local convergence. Furthermore, the complexity of the subalgorithm is not prohibitive. Formulas for $f_\Ubf$ and $f_{\Ubf\Ubf}$ can be precomputed and stored for typical utility functions. The matrix exponential in (\ref{eq:matrixExponential}) involves a matrix with small size $(2M_i) \times (2M_i)$. There exist even simpler alternative ways to generating a curve with a given tangent vector other than the geodesic  \cite{Absil&etal:book,Wen&Yin:10Technical}. The subalgorithm presented here is only one example among many possible implementations for optimization on Stiefel manifolds and a variety of different methods are available to compromise complexity and performance \cite{Absil&etal:book,Manton:02SP}.

\subsection{A Design Example: Weighted Sum Rate Maximization}
\label{subsec:weightedsumrateMax}

In this subsection, we provide a specific example for the proposed beam design method. Here, we consider the cooperative beam design for weighted sum rate maximization by using the steepest descent on the product manifold $\Mc_1\times \cdots \times \Mc_K$. The weighted sum
rate maximization problem is formulated as
\begin{eqnarray}
& & \max_{\{\Ubf_k, \boldsymbol{\Lambda}_k,k=1,\cdots,K\}}~  \sum_{i=1}^K w_i R_i   \\
&=& \min_{\{\Ubf_k, \boldsymbol{\Lambda}_k,k=1,\cdots,K\}}~  - \sum_{i=1}^K w_i \log\Big|
\Ibf+\Big(\Ibf+\sum_{j\neq i} \Hbf_{ij}\Qbf_j\Hbf_{ij}^H\Big)^{-1}\Hbf_{ii}\Qbf_i\Hbf_{ii}^H \Big|, \nonumber
\end{eqnarray}
where $\Qbf_i = \Upsilonbf_i\Ubf_i\Lambdabf_i\Ubf_i^H\Upsilonbf_i^H$ and $\{w_i: w_i \ge 0,~ \sum_i w_i =1\}$ is the set of weighting factors.
To compute the gradient of the objective function,  we manipulate the rate formula of receiver $i$ as follows. First, consider the case of $k=i$.
\begin{eqnarray}
 R_i
&=& \log\Big|
\Ibf+(\Ibf+\sum_{j\neq i} \Hbf_{ij}\Qbf_{j}\Hbf_{ij}^H)^{-1}\Hbf_{ii}\Qbf_i\Hbf_{ii}^H \Big| \nonumber\\
&=& \log\Big|
\Ibf+\sum_{j\neq i} \Hbf_{ij}\Qbf_j\Hbf_{ij}^H+\Hbf_{ii}\Qbf_i\Hbf_{ii}^H \Big|-\log\Big|
\Ibf+\sum_{j\neq i} \Hbf_{ij}\Qbf_j\Hbf_{ij}^H \Big| \nonumber\\
&=& \log\Big|
\underbrace{\Ibf+\sum_{j\neq k}
\Hbf_{ij}\Qbf_j\Hbf_{ij}^H}_{=:\Abf_{kk}\Abf_{kk}^H}+\Hbf_{kk}\Qbf_k\Hbf_{kk}^H \Big|
- \mbox{constant} \nonumber
\end{eqnarray}
\begin{eqnarray}
&=& \log\Big|
\Ibf+\Abf_{kk}^{-1}\Hbf_{kk}\Qbf_k\Hbf_{kk}^H\Abf_{kk}^{-H} \Big|
- \mbox{constant} \nonumber \\
&=& \log\Big|
\Ibf+\underbrace{\Abf_{ik}^{-1}\Hbf_{kk}\Upsilonbf_{k}}_{=:\Cbf_{kk}^H}\Ubf_k\Lambdabf_k\Ubf_k^H \underbrace{\Upsilonbf_{k}^H\Hbf_{kk}^H\Abf_{ik}^{-H}}_{=\Cbf_{kk}} \Big|
- \mbox{constant}  \nonumber\\
&=& \log\Big|
\Ibf+\Ubf_k^H\Cbf_{kk}\Cbf_{kk}^H\Ubf_k\Lambdabf_k\Big|
- \mbox{constant}  \label{eq:Riklambda} \\
&=& \log\Big|
\Lambdabf_k^{-1}+\Ubf_k^H\Cbf_{kk}\Cbf_{kk}^H\Ubf_k\Big|
+\log|\Lambdabf_k|- \mbox{constant}.
\end{eqnarray}
Thus, the (Wirtinger) derivative of $R_i$ w.r.t. $\Ubf_k^*$ for $k=i$ is given by \cite{Kay:book}
\begin{equation}
\frac{\partial R_i}{\partial \Ubf_k^*}
= \Cbf_{kk}\Cbf_{kk}^H\Ubf_k \left( \Lambdabf_k^{-1}+\Ubf_k^H\Cbf_{kk}\Cbf_{kk}^H\Ubf_k\right)^{-1}.
\end{equation}
For the gradient of $R_i$ w.r.t. $\Lambdabf_k$, we only need to consider
the diagonal elements of $\Lambdabf_k=(\lambda_{k1},\cdots,\lambda_{kM_k})$ since the off-diagonal elements are fixed to zero. In (\ref{eq:Riklambda}),  define $\tilde{\Cbf}_{kk} := \Cbf_k^H\Ubf_k$. In the case of $i=k$, from (\ref{eq:Riklambda}),  the gradient of $R_i$ w.r.t. $\lambda_{kl}$ for $l\in\{1,\cdots,M_k\}$ is obtained as
\begin{eqnarray}
\frac{\partial R_k}{\partial \lambda_{kl}}
&=& 
\frac{\partial }{\partial \lambda_{kl}}
\bigg[
\log\Big| \Ibf+ \tilde\Cbf_{kk}\Lambdabf_k\tilde\Cbf_{kk}^H \Big| 
- \mbox{constant} 
\bigg] \nonumber \\
&=&
\mbox{tr}\bigg[
\Big(\Ibf+ \tilde\Cbf_{kk}\Lambdabf_k\tilde\Cbf_{kk}^H\Big)^{-1}
\left(
\frac{\partial}{\partial \lambda_{kl}}
(\Ibf+ \tilde\Cbf_{kk}\Lambdabf_k\tilde\Cbf_{kk}^H)
\right)
\bigg] \nonumber \\
&=&
\mbox{tr}\bigg[
\Big(\Ibf+ \tilde\Cbf_{kk}\Lambdabf_k\tilde\Cbf_{kk}^H\Big)^{-1}
\bigg(
\frac{\partial}{\partial \lambda_{kl}}	
\Big(\Ibf+ \sum_{l=1}^{M_k}\lambda_{kl}\tilde\Cbf_{kk}(:, l)
\tilde\Cbf_{kk}^H(:, l)\Big)
\bigg)
\bigg] \nonumber \\
&=&
\mbox{tr}\bigg[
\Big(\Ibf+ \tilde\Cbf_{kk}\Lambdabf_k\tilde\Cbf_{kk}^H\Big)^{-1}
\Big(\tilde\Cbf_{kk}(:, l)
\tilde\Cbf_{kk}^H(:, l)\Big)
\bigg] \nonumber \\
&=&
\tilde\Cbf_{kk}^H(:, l)
\Big(\Ibf+ \tilde\Cbf_{kk}\Lambdabf_k\tilde\Cbf_{kk}^H\Big)^{-1}
\tilde\Cbf_{kk}(:, l),
\end{eqnarray}
where $\tilde{\Cbf}_{kk}(:,l)$ is the $l$-th column of $\tilde{C}_{kk}$
and the second equality is from \cite{Kay:book}.
Therefore, 
\begin{equation}
\frac{\partial R_k}{\partial \Lambdabf_k}
=\Big[\tilde\Cbf_{kk}^H(:, 1)
\Big(\Ibf+ \tilde\Cbf_{kk}\Lambdabf_k\tilde\Cbf_{kk}^H\Big)^{-1}
\tilde\Cbf_{kk}(:,1),\ldots,
\tilde\Cbf_{kk}^H(:, M_k)
\Big(\Ibf+ \tilde\Cbf_{kk}\Lambdabf_k\tilde\Cbf_{kk}^H\Big)^{-1}
\tilde\Cbf_{kk}(:,M_k)
\Big]^T.
\end{equation}
Next, consider the case of $i \ne k$. In this case,
\begin{eqnarray}
 R_i
&=& \log\Big|
\Ibf+(\Ibf+\sum_{j\neq i} \Hbf_{ij}\Qbf_{j}\Hbf_{ij}^H)^{-1}\Hbf_{ii}\Qbf_i\Hbf_{ii}^H \Big| \nonumber\\
&=& \log\Big|
\Ibf+\sum_{j\neq i} \Hbf_{ij}\Qbf_j\Hbf_{ij}^H+\Hbf_{ii}\Qbf_i\Hbf_{ii}^H \Big|-\log\Big|
\Ibf+\sum_{j\neq i} \Hbf_{ij}\Qbf_j\Hbf_{ij}^H \Big| \nonumber\\
&=& \log\Big|
\underbrace{\Ibf+\sum_{j\neq k}
\Hbf_{ij}\Qbf_j\Hbf_{ij}^H}_{=:\Abf_{ik}\Abf_{ik}^H}+\Hbf_{ik}\Qbf_k\Hbf_{ik}^H \Big|
-\log\Big|
\underbrace{\Ibf+\sum_{j\neq i, j\neq k} \Hbf_{ij}\Qbf_{j}\Hbf_{ij}^H}_{=:\Bbf_{ik}\Bbf_{ik}^H}
+ \Hbf_{ik}\Qbf_{k}\Hbf_{ik}^H \Big| \nonumber\\
&=& \log\Big|
\Ibf+\Abf_{ik}^{-1}\Hbf_{ik}\Qbf_k\Hbf_{jk}^H\Abf_{ik}^{-H} \Big|
-\log\Big|
\Ibf+\Bbf_{ik}^{-1}\Hbf_{ik}\Qbf_{k}\Hbf_{ik}^H\Bbf_{ik}^{-H}\Big|
+ \mbox{constant}\nonumber
\end{eqnarray}
\begin{eqnarray}
&=& \log\Big|
\Ibf+\underbrace{\Abf_{ik}^{-1}\Hbf_{ik}\Upsilonbf_{k}}_{=:\Cbf_{ik}^H}\Ubf_k\Lambdabf_k\Ubf_k^H\underbrace{\Upsilonbf_{k}^H\Hbf_{ik}^H\Abf_{ik}^{-H}}_{=\Cbf_{ik}} \Big|
-\log\Big|
\Ibf+\underbrace{\Bbf_{ik}^{-1}\Hbf_{ik}\Upsilonbf_{k}}_{=:\Dbf_{ik}^H}\Ubf_k\Lambdabf_k\Ubf_k^H\underbrace{\Upsilonbf_{k}^H\Hbf_{ik}^H\Bbf_{ik}^{-H}}_{=\Dbf_{ik}}\Big|
+ \mbox{constant} \nonumber\\
&=& \log\Big|
\Ibf+\Ubf_k^H\Cbf_{ik}\Cbf_{ik}^H\Ubf_k\Lambdabf_k\Big|
-\log\Big|
\Ibf+\Ubf_k^H\Dbf_{ik}\Dbf_{ik}^H\Ubf_k\Lambdabf_k\Big| + \mbox{constant}  \label{eq:Rikdiff1}  \\
&=& \log\Big|
\Lambdabf_k^{-1}+\Ubf_{k}^H\Cbf_{ik}\Cbf_{ik}^H\Ubf_k\Big|
-\log\Big|
\Lambdabf_k^{-1}+\Ubf_k^H\Dbf_{ik}\Dbf_{ik}^H\Ubf_k\Big|
+\log|\Lambdabf_k|-\log|\Lambdabf_k| + \mbox{constant}.  \label{eq:Rikdiff2}
\end{eqnarray}
From (\ref{eq:Rikdiff1}) and (\ref{eq:Rikdiff2}), the derivatives of $R_i$ w.r.t. $\Ubf_k^*$ and $\Lambdabf_k$ are respectively given for $i\ne k$ by
\begin{equation}
\frac{\partial R_i}{\partial \Ubf_k^*}
= 
\Ubf_k^H\Cbf_{ik}\Cbf_{ik}^H\Ubf_k \left( \Ibf + \Ubf_k^H\Cbf_{ik}\Cbf_{kk}^H\Ubf_k \Lambdabf_k \right)^{-1}
-
\Ubf_k^H\Dbf_{ik}\Dbf_{ik}^H\Ubf_k \left( \Ibf + \Ubf_k^H\Dbf_{ik}\Dbf_{ik}^H\Ubf_k \Lambdabf_k \right)^{-1}
\end{equation}
and
\begin{eqnarray}
\frac{\partial R_i}{\partial \Lambdabf_k}
&=& \Big[
\tilde\Cbf_{ik}^H(:, 1)
\Big(\Ibf+ \tilde\Cbf_{ik}\Lambdabf_k\tilde\Cbf_{ik}^H\Big)^{-1}
\tilde\Cbf_{ik}(:,1)
-
\tilde\Dbf_{ik}^H(:, 1)
\Big(\Ibf+ \tilde\Dbf_{ik}\Lambdabf_k\tilde\Dbf_{ik}^H\Big)^{-1}
\tilde\Dbf_{ik}(:,1),
\cdots, \\
& & 
\tilde\Cbf_{ik}^H(:, M_k)
\Big(\Ibf+ \tilde\Cbf_{ik}\Lambdabf_k\tilde\Cbf_{ik}^H\Big)^{-1}
\tilde\Cbf_{ik}(:,M_k)
-
\tilde\Dbf_{ik}^H(:, M_k)
\Big(\Ibf+ \tilde\Dbf_{ik}\Lambdabf_k\tilde\Dbf_{ik}^H\Big)^{-1}
\tilde\Dbf_{ik}(:,M_k),
\Big]^T, \nonumber
\end{eqnarray}
where $\tilde{\Cbf}_{ik} = \Cbf_{ik}^H\Ubf_k$ and $\tilde{\Dbf}_{ik}=\Dbf_{ik}^H\Ubf_k$.
Then, the derivatives of the overall cost function w.r.t. $\Ubf_k^*$ and $\Lambdabf_k$ are given respectively by
\begin{equation}\label{eq:wsr_gradient}
\frac{\partial }{\partial \Ubf_k^*} \Big( \sum_{i=1}^K w_i R_i \Big)
=
\sum_{i=1}^K w_i \frac{\partial R_i}{\partial \Ubf_k^*}
\end{equation}
and
\begin{equation}\label{eq:wsr_gradient_power}
\frac{\partial }{\partial \Lambdabf_k} \Big( \sum_{i=1}^K w_i R_i \Big)
=
\sum_{i=1}^K w_i \frac{\partial R_i}{\partial \Lambdabf_k}
\end{equation}
for $k=1,\cdots,K$. With the obtained derivatives, Algorithm \ref{algo:overall} with the subalgorithm, Algorithm \ref{algo:newton}, can be run.

\section{Numerical Results}
\label{sec:numerical}

In this section, we provide some numerical results to validate our beam design paradigm based on the parameterization $\{\Mc_i\}$ for the beam search space for MIMO interference channels. We here consider the weighted sum rate maximization problem proposed in Section \ref{subsec:weightedsumrateMax}.

\begin{figure}[ht]
\centering
\includegraphics[width=3.8in]{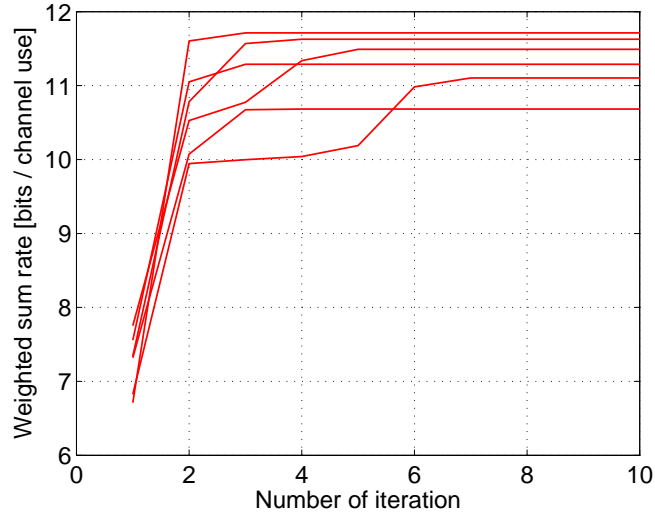}
\caption{Convergence of Algorithm \ref{algo:overall}: $K=3$, $(N,M_1,M_2, M_3)=(8,2,2,2)$, and $P_1=P_2=P_3=30$.}
\label{fig:convergence}
\end{figure}
First, we verified the convergence of the overall algorithm. Fig. \ref{fig:convergence} shows the convergence behavior of Algorithm \ref{algo:overall} for several different channel realizations when $K=3$, $(N,M_1,M_2, M_3)=(8,2,2,2)$ and $P_1=P_2=P_3=30$. Here, we used the steepest descent method on the product manifold $\Mc_i$ with step sizes 
\begin{equation}\label{eq:step_sizes}
0.05 \times \bigg\|\frac{\partial}{\partial \Ubf_k^*} 
\bigg(\sum_{i=1}^K w_i R_i\bigg)  \bigg\|_F
~~~~ \mbox{and} ~~~~
0.05 \times \bigg\|\frac{\partial}{\partial \Lambdabf_k} 
\bigg(\sum_{i=1}^K w_i R_i\bigg)  \bigg\|_F
\end{equation}
for the $U$ and $\Lambda$ steps in our subalgorithm. 
The step sizes in \eqref{eq:step_sizes} are designed to gradually reduce to zero as the subalgorithm 
approaches a (locally) optimal point and to show better convergence behavior near
the locally optimal  point. 
It is observed in the figure that the overall algorithm converges very fast and the number of iterations for convergence is only a few for most channel realizations in this case. Thus, the main computational time lies in the execution of the subalgorithm.  Although the steepest descent based subalgorithm is used  in this demonstration,  different methods with faster convergence can be used \cite{Absil&etal:book,Wen&Yin:10Technical,Manton:02SP}.

\begin{figure*}[h]
\centerline{
\SetLabels
\L(0.25*-0.1) (a) \\
\L(0.75*-0.1) (b) \\
\endSetLabels
\leavevmode
\strut\AffixLabels{
\scalefig{0.5}\epsfbox{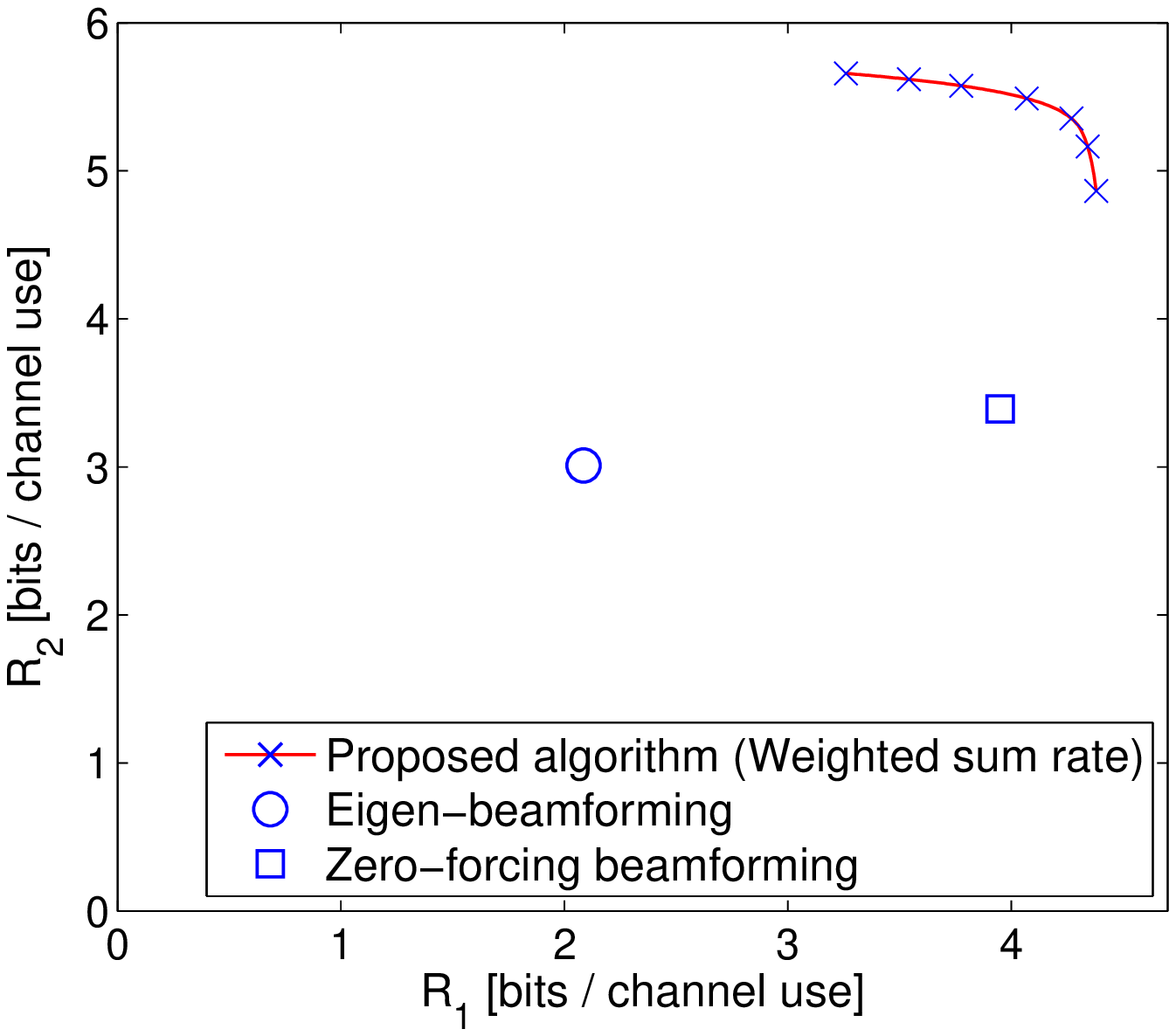}
\scalefig{0.5}\epsfbox{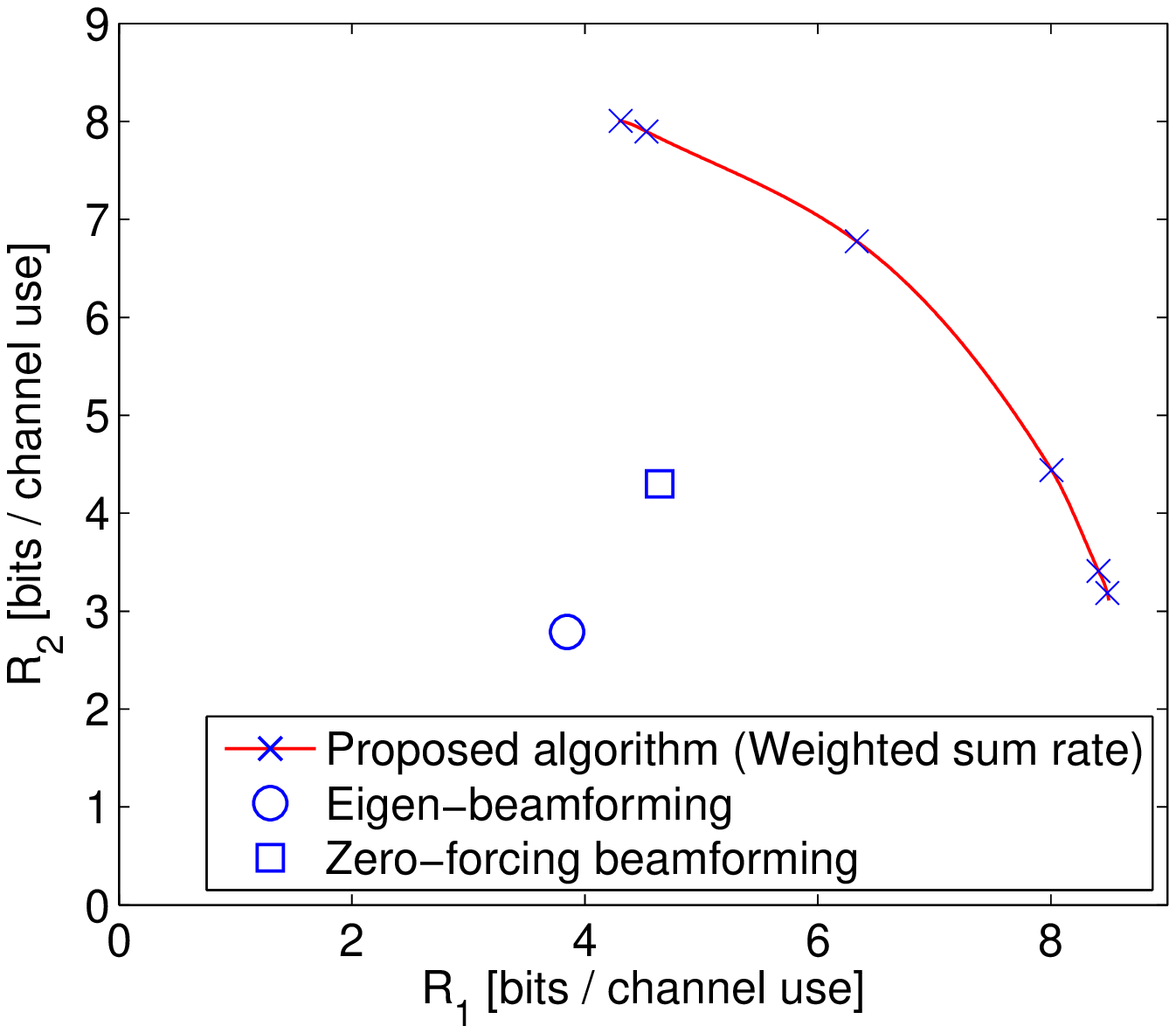} } }
\vspace{1.5em} \caption{Rate pairs of several beam design methods: (a) $K=2$, $(N,M_1,M_2)=(6, 2, 2)$, $P_1=P_2=5$ and (b) $K=2$, $(M,N_1, N_2)=(5, 2, 2)$, $P_1=P_2=10$} \label{fig:Pareto}
\end{figure*}

 With convergence of the algorithm confirmed, we examined the sum rate performance of the proposed beam design algorithm. Figures \ref{fig:Pareto} (a) and (b) show the rate-tuples of several beam design methods for two different channel  settings.  We considered the single-user eigen-beamforming, the zero-forcing beamforming in addition to the proposed beam design method. For the proposed beam design method for weighted sum rate maximization, we varied the weights so that we can obtain rate-tuples at different locations. As expected, it is seen that the rate performance of the proposed method is superior to those of  the eigen-beamforming and the zero-forcing. Of course, the weighted sum rate maximization can be performed in the original beam search space $\Qc_1 \times \cdots \times \Qc_K$ by using one of gradient descent type algorithms. However, such a method is far less efficient than the proposed beam design method based on the proposed parameterization for the beam search space not losing Pareto-optimality.

\begin{figure}[h]
\centering
\includegraphics[width=4in]{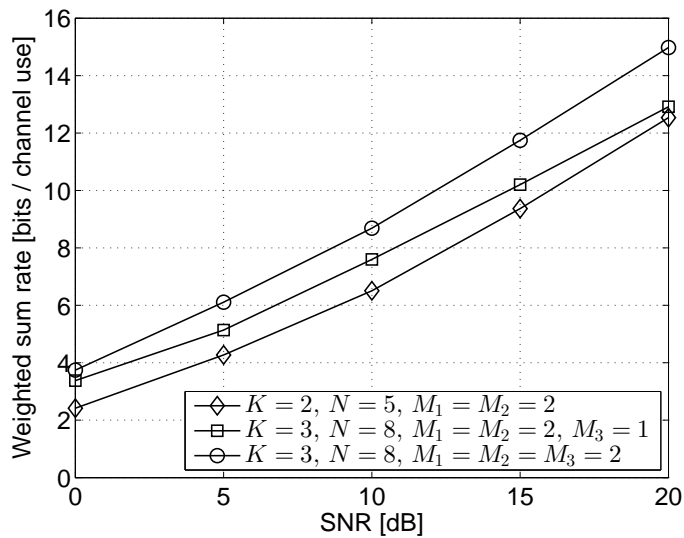}
\caption{Performance of the proposed method: Sum rate with respect to SNR}
\label{fig:rate_vs_snr}
\end{figure}

 Finally, 
Fig. \ref{fig:rate_vs_snr} shows the  sum rate performance of the algorithm w.r.t. SNR for three different system parameter settings: (1) $K=2$, $(N,M_1,M_2)=(5,2,2)$, (2) $K=3$, $(N,M_1,M_2, M_3)=(8,2,2,1)$, and (3) $K=3$, $(N,M_1,M_2, M_3)=(8,2,2,2)$.  Table \ref{table:DoF} summarizes the corresponding obtained rank of the converged $\Lambdabf_i$, $i=1,\cdots,K$, by the proposed beam design algorithm. Note that in the low SNR regime indeed the proposed beam design algorithm does not yield a beamformer with the maximum available DoFs of $M_i$ for all $i$. Futhermore, it tells who should not use the available (single-user) full DoFs  for sum rate maximization.  It is expected that at low SNR  the optimal strategy does not use maximum DoFs since all power can be allocated in the best direction, as in the single-user MIMO case. Due to the separate parameterization for $\Ubf_i$ and $\Lambdabf_i$ in the beam search space $\Mc_i$, our algorithm can clearly identify the optimal rank of the beamforming matrix  by checking the rank of $\Lambdabf_i$.

\begin{table}[h] 
\begin{center}
\begin{tabular}{|c|c|c|c|c|c|} 
\hline
SNR [dB]  & 0 & 5 & 10 & 15 & 20  \\ \hline
$K=2$, $N=5, M_1=2, M_2=2$  & $(1,2)$ & $(1,2)$ & $(1,2)$ & $(2,2)$ & $(2,2)$  \\ \hline
$K=3$, $N=8, M_1=2, M_2=2, M_3=1$  & $(1,2,1)$ & $(1,2,1)$ & $(2,2,1)$ & $(2, 2, 1)$ & $(2,2,1)$  \\ \hline
$K=3$, $N=8, M_1=2, M_2=2, M_3=2$  & $(2,1,1)$ & $(2,2,1)$ & $(2,2,2)$ & $(2,2,2)$ & $(2,2,2)$  \\ \hline
\end{tabular}
\end{center}
\caption{The obtained number of data streams for Fig. \ref{fig:rate_vs_snr}: $d_i$ in
$(d_1, d_2, d_3)$ in the table denotes  the obtained rank of $\Lambdabf_i$ for transmitter-receiver pair $i$}
\label{table:DoF}
\end{table}

\section{Conclusion}
\label{sec:conclusion}

We have considered the Pareto-optimal beam structure for multi-user multiple-input multiple-output (MIMO) interference channels and have provided
a necessary condition for any Pareto-optimal transmit signal covariance matrix for  the $K$-pair Gaussian $(N,M_1,\cdots,M_K)$ interference channel. We have shown  that any Pareto-optimal transmit signal covariance matrix at a transmitter should have its column space contained in the union of the eigen-spaces of the channel matrices from the transmitter to all receivers. Based on this necessary condition, we have proposed an efficient parameterization for the beam search space, given by the product manifold of a Stiefel manifold and a subset of a hyperplane. Based on the proposed parameterization, we have developed a very efficient beam design algorithm by exploiting the  geometrical structure of the beam search space and existing tools for optimization on Stiefel manifolds.  We hope that the results in this paper would be helpful for efficient intercell interference control based on MIMO antenna technologies in current and future cellular networks.


\end{document}
